\documentclass[11pt,a4paper]{article}

\usepackage{indentfirst}
\usepackage{graphicx}
\usepackage{epsfig}

\usepackage{latexsym}
\usepackage{amsmath}
\usepackage{amssymb}
\usepackage{amsfonts}
\usepackage{mathrsfs}

\usepackage{morefloats}

\usepackage{natbib}

\newtheorem{lemma}{Lamma}
\newtheorem{proposition}{Proposition}

\newcommand{\mbf}{\boldsymbol}

\newcommand{\bnabla}{\boldsymbol{\nabla}}
\newcommand{\bnb}{\boldsymbol{\nabla}}

\newcommand{\ts}{\times}
\newcommand{\ots}{\otimes}
\newcommand{\bcdot}{\boldsymbol{\cdot}}
\newcommand{\bdot}{\boldsymbol{\cdot}}

\newcommand{\pl}{\partial}
\newcommand{\eps}{\varepsilon}
\newcommand{\nb}{\nabla}

\newcommand{\os}{\overset}
\newcommand{\ol}{\overline}

\newcommand{\s}{\Sigma}

\begin{document}

\begin{center}
{
    \Large\textbf{On Two Kinds of Differential Operators on General Smooth Surfaces}
    \footnote{Corresponding Author: XIE XiLin, Department of Mechanics \& Engineering Science, Fudan University.
              HanDan Road 220, Shanghai 200433, China. Tel: 0086-21-55664283; Email: xiexilin@fudan.edu.cn}
}
\end{center}

\begin{center}
{   \small
    XIE XiLin\\
    Department of Mechanics \& Engineering Science, Fudan University,\\
    Shanghai 200433, China.\\
}
\end{center}

\begin{center} \textbf{Original Manuscript updated on {\today} }\end{center}

\begin{abstract}
    Two kinds of differential operators that can be generally defined on an arbitrary smooth surface in a finite
    dimensional Euclid space are studied, one is termed as surface gradient and the other one as Levi-Civita gradient.
    The surface gradient operator is originated from the differentiability of a tensor field defined on the surface.
    Some integral and differential identities have been theoretically studied that play the important role in the studies
     on continuous mediums whose geometrical configurations can be taken as surfaces and on interactions between fluids
     and deformable boundaries. The definition of Levi-Civita gradient is based on Levi-Civita connections generally defined
     on Riemann manifolds. It can be used to set up some differential identities in the intrinsic/coordiantes-independent
     form that play the essential role in the theory of vorticity dynamics for two dimensional flows on general fixed
     smooth surfaces.
\end{abstract}


\section{Introduction}

\subsection{Fundamentals of differential calculus on a surface}
Generally, an $m$-dimensional surface in $m+1$ Euclid space can be represented as
\begin{equation*}
    \mbf{\Sigma}(x_\Sigma,t):\,
    \mathbb{R}^m\supset\mathscr{D}_x\ni\mbf{x}_\Sigma\,\mapsto\,\mbf{\Sigma}(x_\Sigma,t)\in\mathbb{R}^{m+1}
\end{equation*}
In the case that $\mbf{x}_\Sigma$ is a nonsingular point, $\{\mbf{g}_i(x_\Sigma,t):=\frac{\pl\mbf{\Sigma}}{\pl
x^i_\Sigma}(x_\Sigma,t)\}^m_{i=1}$ constitutes the so-called covariant basis of the tangent space
$\mbf{T}_x\mbf{\Sigma}$ and there exists uniquely one direction $\mbf{n}(x_\Sigma,t)$ that is particular to the
tangent space, i.e. $(\mbf{n},\mbf{g}_i)(x_\Sigma,t)_{\mathbb{R}^{m+1}}=0\,(i=1,\cdots,m)$.

Two kinds of the fundamental affine tensor could be defined
\begin{alignat*}{3}
&   \mbf{G}\triangleq g_{ij}\mbf{g}^i\otimes\mbf{g}^j,\quad
&&  g_{ij}:=(\mbf{g}_i,\mbf{g}_j)(x_\Sigma,t)_{\mathbb{R}^{m+1}}\\
&   \mbf{K}\triangleq b_{ij}\mbf{g}^i\otimes\mbf{g}^j,\quad
&&  b_{ij}:=\left( \frac{\pl\mbf{g}_i}{\pl x^j_\Sigma}(x_\Sigma,t),\mbf{n} \right)_{\mathbb{R}^{m+1}}
\end{alignat*}
that are termed as the metric tensor and the curvature tensor respectively. Gaussian curvature is defined as $K_G:=\det[b_{ij}]/\det[g_{ij}]=\det[b^i_j]=:\det\mbf{B}$ and mean curvature as $H:=b^s_s =:tr\mbf{K}$.

Based on the differential calculus in $\mathbb{R}^{m+1}$, one has the following so-termed frame movement equations
\begin{equation*}
\left\{
\begin{array}{l}
\frac{\pl\mbf{g}_i}{\pl x^j_\Sigma}(x_\Sigma,t)= \Gamma^k_{ji}\mbf{g}_k+b_{ji}\mbf{n}=
 \Gamma_{ji,k}\mbf{g}^k+b_{ji}\mbf{n}\\
\frac{\pl\mbf{g}^i}{\pl x^j_\Sigma}(x_\Sigma,t)=-\Gamma^i_{jk}\mbf{g}^k+b^i_j\mbf{n}
\end{array}
\right.;\quad
\frac{\pl\mbf{n}}{\pl x^j_\Sigma}(x_\Sigma,t)=-b_{jk}\mbf{g}^k=-b_j^k\mbf{g}_k
\end{equation*}
where $\Gamma_{ji,k}$ and $\Gamma^k_{ji}$ are the Christoffel symbols of the first and second kinds respectively. In addition, one has the relation between metric tensor and Christoffel symbol
\begin{equation*}
    \Gamma_{ij,k}=\frac{1}{2}\left( \frac{\pl g_{jk}}{\pl x^i_\s}+ \frac{\pl g_{ik}}{\pl x^j_\s}
    - \frac{\pl g_{ij}}{\pl x^k_\s}\right)(x_\s,t)
\end{equation*}

An $m$ dimensional smooth surface embedded in $m+1$ dimensional Euclid space is naturally a Riemann manifold with the metric represented by the metric tensor and the covariant derivative/differentiation denoted by $\nb_l$ defined as, say $\mbf{\Phi}:=\Phi^{ij}_{\cdot\cdot k}\mbf{g}_i\ots\mbf{g}_j\ots\mbf{g}^k\in\mathscr{T}^3(\mbf{T\Sigma})$ is a tensor field with order $3$ on the surface,
\begin{equation*}
    \nb_l\Phi^{ij}_{\cdot\cdot k}
   \triangleq \frac{\pl \Phi^{ij}_{\cdot\cdot k} }{\pl x^l_\s}(x_\s,t)+\Gamma^i_{ls}\Phi^{sj}_{\cdot\cdot k}
   +\Gamma^j_{ls}\Phi^{is}_{\cdot\cdot k}-\Gamma^s_{lk}\Phi^{ij}_{\cdot\cdot s}
\end{equation*}

The fundamentals of differential calculus on a surface can be referred to the monographs by \cite{Durovin-1992} and
\cite{GuoZH-1980}.

\subsection{Sketch of the present paper}

Two kinds of differential operators on the surface are to be studied that are termed as surface gradient operator and Levi-civita gradient operator respectively. The whole content of the present paper can be divided into two parts. The first part is on the surface gradient tensor that is originated from the differentiation of a tensor field defined on the surface. As applications, four related aspects in fluid and solid mechanics are referred that include $\S\,2.1$ intrinsic generalized Stokes formulas with three kinds of applications, $\S\,2.2$ primary properties of deformation gradient tensor for thin enough continuous mediums, $\S\,2.3$ strain tensor on an arbitrary deformable surface. The second part is on the Levi-Civita gradient operator that is based on Levi-Civita connection possessed by any Riemann manifold. Its applications refer to $\S\,3.1$ some primary identities in vorticity dynamics of two dimensional flows on fixed smooth surfaces and $\S\,3.2$ some identities of affine surface tensors.

Generally, the surface gradient operator is more familiar to mechanicians and Levi-Civita connection is to mathematicians. However, all of the applications as indicated in the present paper are closely linked to the mechanics of continuous mediums whose geometrical configurations are either bulks or surfaces. And all of the related results accompanying with deductions are independent to others studies.

\section{Surface gradient operator}

Generally, the \emph{surface gradient operator} $\overset{\Sigma}{\bnabla}\equiv \boldsymbol{g}^l\frac{\partial}{\partial x^l_\Sigma}$ is defined as, say $\mathbf{\Phi}\in\mathscr{T}^2(\mathbb{R}^m)$,
\begin{alignat*}{3}
    \overset{\Sigma}{\bnabla}\circ-\mathbf{\Phi}
&   \equiv\left(\boldsymbol{g}^l\frac{\partial}{\partial x^l_\Sigma}\right)\circ-
    \left( \Phi^i_{\cdot j}\boldsymbol{g}_i\otimes\boldsymbol{g}^j +\Phi^i_{\cdot 3}\mbf{g}_i\otimes\mbf{n}+\Phi^3_{\cdot j}\mbf{n}\otimes\mbf{g}^j+\Phi^3_{\cdot 3}\mbf{n}\otimes\mbf{n} \right)\\
&   \triangleq \boldsymbol{g}^l\circ-
    \frac{\partial}{\partial x^l_\Sigma}\left( \Phi^i_{\cdot j}\boldsymbol{g}_i\otimes\boldsymbol{g}^j +\Phi^i_{\cdot 3}\mbf{g}_i\otimes\mbf{n}+\Phi^3_{\cdot j}\mbf{n}\otimes\mbf{g}^j+\Phi^3_{\cdot 3}\mbf{n}\otimes\mbf{n} \right)\\
&   =
    \left[ \nabla_l\Phi^i_{\cdot j}(\boldsymbol{g}^l\circ-\boldsymbol{g}_i)\otimes\boldsymbol{g}^j
    +\Phi^i_{\cdot j}b_{li}(\boldsymbol{g}^l\circ-\boldsymbol{n})\otimes\boldsymbol{g}^j
    +\Phi^i_{\cdot j}b_l^j(\boldsymbol{g}^l\circ-\boldsymbol{g}_i)\otimes\boldsymbol{n} \right]\\
&   +
    \left[ \nabla_l\Phi^i_{\cdot 3}(\boldsymbol{g}^l\circ-\boldsymbol{g}_i)\otimes\boldsymbol{n}
    +\Phi^i_{\cdot 3}b_{li}(\boldsymbol{g}^l\circ-\boldsymbol{n})\otimes\boldsymbol{n}
    -\Phi^i_{\cdot 3}b_l^s(\boldsymbol{g}^l\circ-\boldsymbol{g}_i)\otimes\boldsymbol{g}_s \right]\\
&   +
    \left[ \nabla_l\Phi^3_{\cdot j}(\boldsymbol{g}^l\circ-\boldsymbol{n})\otimes\boldsymbol{g}^j
    -\Phi^3_{\cdot j}b_l^s(\boldsymbol{g}^l\circ-\boldsymbol{g}_s)\otimes\boldsymbol{g}^j
    +\Phi^3_{\cdot j}b_l^j(\boldsymbol{g}^l\circ-\boldsymbol{n})\otimes\boldsymbol{n} \right]\\
&   +
    \left[ \nabla_l\Phi^3_{\cdot 3}(\boldsymbol{g}^l\circ-\boldsymbol{n})\otimes\boldsymbol{n}
    -\Phi^3_{\cdot 3}b_l^s(\boldsymbol{g}^l\circ-\boldsymbol{g}_s)\otimes\boldsymbol{n}
    -\Phi^3_{\cdot 3}b_l^s(\boldsymbol{g}^l\circ-\boldsymbol{n})\otimes\boldsymbol{g}_s \right]
\end{alignat*}
where $\circ-$ represents any available algebra tensor operator, $\nabla_l$ denotes the co-variant derivative/differentation of the tensor component that is just effective to the indices with respect to the tangent plane, i.e. $i$, $j$ in the above representations,
\begin{alignat*}{3}
    \nabla_l\Phi^i_{\cdot j}
&\triangleq
    \frac{\partial\Phi^i_{\cdot j}}{\partial x^l_\Sigma}(x_\Sigma,t)+\Gamma^i_{ls}\Phi^s_{\cdot j}
    -\Gamma^s_{lj}\Phi^i_{\cdot s}\\
   \nb_l\Phi^i_{\cdot 3}
&   \triangleq \frac{\pl\Phi^i_{\cdot 3}}{\pl x^l_\Sigma}(x_\Sigma,t)+\Gamma^i_{ls}\Phi^s_{\cdot 3},\quad
    \nb_l\Phi^3_{\cdot j}
    \triangleq \frac{\pl\Phi^3_{\cdot j}}{\pl x^l_\Sigma}(x_\Sigma,t)-\Gamma^s_{lj}\Phi^3_{\cdot s}\\
    \nb_l\Phi^3_{\cdot 3}
&   \triangleq \frac{\pl\Phi^3_{\cdot 3}}{\pl x^l_\Sigma}(x_\Sigma,t)
\end{alignat*}
where $\Gamma^i_{ls}$ denotes Christoffel symbol of the second kind. The contra-variant derivative relates generally to the co-variant one through $\nabla^l\triangleq g^{lt}\nabla_t$. The change of the order of co- and contra-variant derivatives must be related to Riemannian-Christoffel tensor, that is,
\begin{equation*}
    \nabla_p\nabla^q\Phi^i_{\cdot j}
=\nabla^q\nabla_p\Phi^i_{\cdot j}+R^{i\,\cdot\cdot\,q}_{\cdot sp\,\cdot}\Phi^s_{\cdot j}+R^{\cdot s\cdot q}_{j\cdot p\,\cdot}\Phi^i_{\cdot s}
\end{equation*}
where $R^{i\,\cdot\cdot\,q}_{\cdot sp\,\cdot}\triangleq b^i_p b^q_s-b_{sp}b^{iq}$ denotes the component of Riemannian-Christoffel tensor \cite[see][]{GuoZH-1980}. In addition, in the case of two dimensional Riemannian manifolds, Riemannian-Christoffel tensor can be represented by Gaussian curvature and metric tensor as revealed by the relation $R^{i\,\cdot\cdot\,q}_{\cdot sp\,\cdot}=K_G(\delta^i_p \delta^q_s-g_{sp}g^{iq})$.

It should be noted that the definition of the surface gradient operator is based on the differential calculus in the normed linear tensor space, namely, one has
\begin{equation*}
    \mbf{\Phi}(x_\Sigma+\Delta x_\Sigma,t)-\mbf{\Phi}(x_\Sigma,t)
   =\left\{ \begin{array}{c}
   \left( \Delta x^s_\Sigma\mbf{g}_s \right)\bdot\left(\os{\Sigma}{\bnb}\otimes\mbf{\Phi} \right)\\
   \left(\mbf{\Phi}\otimes\os{\Sigma}{\bnb}\right)\bdot \left( \Delta x^s_\Sigma\mbf{g}_s \right)
   \end{array}
   \right.+\mbf{o}(\Delta x)
\end{equation*}
Say $\mbf{\Phi}=\Phi^{i}_{\bdot j 3}\mbf{g}_i\ots\mbf{g}^j\ots\mbf{n}\in\mathscr{T}^3(\mathbb{R}^3)$, one has
\begin{equation*}
    \mbf{\Phi}(x_\s+\Delta x_\s,t)
   =\Phi^{i}_{\bdot j 3}(x_\s+\Delta x_\s,t)(\mbf{g}_i\ots\mbf{g}^j\ots\mbf{n})(x_\s+\Delta x_\s,t)\in\mathscr{T}^3(\mathbb{R}^3)
\end{equation*}
with the differentiations of the tensor component and basis vectors
\begin{alignat*}{3}
& \Phi^{i}_{\bdot j 3}(x_\s+\Delta x_\s,t)=\Phi^{i}_{\bdot j 3}(x_\s,t)+\frac{\pl \Phi^{i}_{\bdot j 3} }{\pl x^s_\s}(x_\s,t)\Delta x^s_\s+o^{i}_{\bdot j 3}(\Delta x_\s)\in\mathbb{R}\\
& \mbf{g}_i(x_\s+\Delta x_\s,t)=\mbf{g}_i(x_\s,t)+\frac{\pl\mbf{g}_i}{\pl x^s_\s}(x_\s,t)\Delta x^s_\s+\mbf{o}_i(\Delta x_\s)\in\mathbb{R}^{m+1}\\
& \mbf{g}^j(x_\s+\Delta x_\s,t)=\mbf{g}^j(x_\s,t)+\frac{\pl\mbf{g}^j}{\pl x^s_\s}(x_\s,t)\Delta x^s_\s+\mbf{o}^j(\Delta x_\s)\in\mathbb{R}^{m+1}\\
& \mbf{n}(x_\s+\Delta x_\s,t)=\mbf{n}(x_\s,t)+\frac{\pl\mbf{n}}{\pl x^s_\s}(x_\s,t)\Delta x^s_\s+\mbf{o}^3(\Delta x_\s)
\in\mathbb{R}^{m+1}
\end{alignat*}
Accompanying the multi-linearity of the representation of any simple tensor with the frame movement equations, the above mentioned representation can be attained. In the view of differentiation, the full dimensional gradient of a tensor filed defined on a domain can be taken as its derivative \cite[see][]{XXL-2012}. Similarly, the surface gradient of a tensor field defined on a surface is its derivative also.

Consequently, the partial derivative of the tensor with respect to one of the component of the surface coordinates can be determined
\begin{equation*}
    \frac{\pl\mbf{\Phi}}{\pl x^l_\Sigma}(x_\Sigma,t)
    \triangleq \lim_{\lambda\rightarrow 0}\,\frac{\mbf{\Phi}(x_\Sigma+\lambda\mathbf{i}_l,t)-\mbf{\Phi}(x_\Sigma,t)}{\lambda}
    =\mbf{g}_l\bdot\left(\os{\Sigma}{\bnb}\otimes\mbf{\Phi}\right)
    =\left( \mbf{\Phi}\otimes \os{\Sigma}{\bnb} \right)\bdot\mbf{g}_l
\end{equation*}
due to
\begin{alignat*}{3}
    \mbf{\Phi}(x_\Sigma+\lambda\mathbf{i}_l,t)-\mbf{\Phi}(x_\Sigma,t)
   =\left\{
   \begin{array}{l}
        (\lambda\mbf{g}_l)\bdot\left(\os{\Sigma}{\bnb}\otimes\mbf{\Phi}\right)+\mbf{o}(\lambda)\\
        \left( \mbf{\Phi}\otimes \os{\Sigma}{\bnb} \right)\bdot(\lambda\mbf{g}_l)+\mbf{o}(\lambda)
   \end{array}
   \right.
\end{alignat*}

\subsection{Intrinsic generalized Stokes formulas}

\begin{proposition}[Generalized Stokes formulas of the first kind]
\begin{alignat*}{3}
    \oint_{C}\,\mbf{\tau}\circ-\mbf{\Phi}\,dl
&  =\int_\Sigma\, \left( \mbf{n}\times\os{\Sigma}{\mbf{\nabla}} \right)\circ-\mbf{\Phi}\,d\sigma\\
    \oint_{C}\,\mbf{\Phi}\circ-\mbf{\tau}\,dl
&  =\int_\Sigma\, \mbf{\Phi}\circ-\left( \mbf{n}\times\os{\Sigma}{\mbf{\nabla}} \right) \,d\sigma
\end{alignat*}
\end{proposition}
\noindent\textbf{Proof:}\quad

It is well known that Stokes formula in the fundamental calculus takes the following form
\begin{equation*}
    \oint_{\pl\Sigma}\,\mbf{\tau}\bdot\mbf{a}\,dl=\int_\Sigma\,\mbf{n}\bdot(\bnb\ts\mbf{a})\,d\sigma\quad
    \mbox{i.e.}\quad
    \oint_{\pl\Sigma}\,\tau_\alpha a_\alpha\,dl=\int_\Sigma\,n_\lambda e_{\lambda\mu\alpha}\frac{\pl a_\alpha}{\pl X_\mu}\,d\sigma
\end{equation*}
where all of the quantities are represented through the canonical basis. Consequently, the vector field $\mbf{a}$
should be extended differentially to a three dimensional open set in which the surface is embedded in order to fulfil
the full dimensional curl operator. This kind of Stokes formula is termed as the prototype in the present paper.

In order to proof the second identity, firstly the integrant of the curve integral is expanded through the canonical
basis, that is
\begin{equation*}
\mathbf{\Phi}\circ-\mbf{\tau}
=(\Phi_{\xi\eta}\,\mathbf{i}_\xi\otimes\mathbf{i}_\eta)\circ-(\tau_\alpha\mathbf{i}_\alpha)
 =\tau_\beta\,\delta_{\beta\alpha}\Phi_{\xi\eta}\,(\mathbf{i}_\xi\otimes\mathbf{i}_\eta\circ-\mathbf{i}_\alpha)
\end{equation*}

Secondly, the stokes formula in the prototype is adopted to attain the surface gradient
\begin{alignat*}{3}
& n_\theta e_{\theta\lambda\beta}\frac{\partial}{\partial
  X^\lambda}(\delta_{\beta\alpha}\Phi_{\xi\eta})\,
  (\mathbf{i}_\xi\otimes\mathbf{i}_\eta\circ-\mathbf{i}_\alpha)
 =n_\theta e_{\theta\lambda\alpha}\frac{\partial\Phi_{\xi\eta}}{\partial
  X^\lambda}\,\mathbf{i}_\xi\otimes\mathbf{i}_\eta\circ-\mathbf{i}_\alpha
 =n_\theta e_{\theta\lambda\alpha}\frac{\pl \mbf{\Phi}}{\pl X^\lambda}\circ-\mathbf{i}_\alpha\\
&=\frac{\pl \mbf{\Phi}}{\pl X^\lambda}\circ-(\mbf{n}\ts\mathbf{i}_\lambda)
 =:\mbf{\Phi}\circ-\left[ \mbf{n}\ts\left( \mathbf{i}_\lambda\frac{\pl}{\pl X^\lambda}\right)\right]
 =:\mbf{\Phi}\circ-(\mbf{n}\ts\bnb)
\end{alignat*}

Thirdly, the full dimensional gradient is represented through the surface gradient
\begin{equation*}
    \mbf{\Phi}\circ-(\mbf{n}\ts\bnb)=\mbf{\Phi}\circ-\left[ \mbf{n}\ts\left(\os{\Sigma}{\bnb}+\mbf{n}\frac{\pl}{\pl X^3} \right) \right]=\mbf{\Phi}\circ-\left( \mbf{n}\ts\os{\Sigma}{\bnb} \right)
\end{equation*}
where the semi-orthogonal curvilinear coordinates \cite[see][]{XXL-2013} is adopted that is a kind of full dimensional curvilinear coordinates. The proof is completed and the second identity can be proved in the same way.

It should be pointed out that the surface gradients are nothing to do with the directional derivative with respect to
the normal direction, in other words the quantity originally defined on the surface does not need to be extended if
the surface gradient rather than the full dimensional gradient is adopted.

\begin{proposition}[Generalized Stokes formulas of the second kind]
\begin{alignat*}{3}
  \oint_C\, (\mbf{\tau}\times\mbf{n})\circ-\,\mbf{\Phi}\,dl
&=\int_\Sigma\,\left( \os{\Sigma}{\mbf{\nabla}}\circ-\mbf{\Phi} + H\mbf{n}\circ-\mbf{\Phi} \right) \,d\sigma\\
  \oint_C\, \mbf{\Phi}\circ-(\mbf{\tau}\times\mbf{n}) \,dl
&=\int_\Sigma\, \left( \mbf{\Phi}\circ-\os{\Sigma}{\mbf{\nabla}} + H\mbf{\Phi}\circ-\mbf{n} \right) \,d\sigma
\end{alignat*}
\end{proposition}

\noindent\textbf{Proof}\quad

The proof of the second identity is carried out as follows. And the first one can be verified in the same way \cite[see][]{XXL-2013}.

Firstly, the integrant of the curve integral is expanded through the canonical basis
\begin{equation*}
\mathbf{\Phi}\circ-(\mbf{\tau}\times\mathbf{n})
=(\Phi_{\xi\eta}\,\mathbf{i}_\xi\otimes\mathbf{i}_\eta)\circ-(e_{\alpha\beta\gamma}\,\tau_\beta\,n_\gamma\,\mathbf{i}_\alpha)
 =\tau_\beta\,e_{\beta\gamma\alpha}n_\gamma\Phi_{\xi\eta}\,
(\mathbf{i}_\xi\otimes\mathbf{i}_\eta\circ-\mathbf{i}_\alpha)
\end{equation*}

Secondly, the curve integral is transferred to the surface integral according to the Stokes formula in the prototype. The deduction of the surface integrant is as follows
\begin{alignat*}{3}
& n_\theta e_{\theta\lambda\beta}\frac{\partial}{\partial
  X^\lambda}(e_{\beta\gamma\alpha}n_\gamma\Phi_{\xi\eta})\,
  (\mathbf{i}_\xi\otimes\mathbf{i}_\eta\circ-\mathbf{i}_\alpha)
 =e_{\theta\lambda\beta}e_{\gamma\alpha\beta}\,n_\theta\,\frac{\partial}{\partial
  X^\lambda}(n_\gamma\Phi_{\xi\eta})\,
  (\mathbf{i}_\xi\otimes\mathbf{i}_\eta\circ-\mathbf{i}_\alpha)\\
&=(\delta_{\theta\gamma}\delta_{\lambda\alpha}-\delta_{\lambda\gamma}\delta_{\theta\alpha})\,n_\theta\,\frac{\partial}{\partial
  X^\lambda}(n_\gamma\Phi_{\xi\eta})\,
  (\mathbf{i}_\xi\otimes\mathbf{i}_\eta\circ-\mathbf{i}_\alpha)\\
&=\left[ \frac{\partial \Phi_{\xi\eta}}{\partial X^\alpha}-n_\alpha\frac{\partial n_\lambda}{\partial
  X^\lambda}\Phi_{\xi\eta}-n_\alpha n_\lambda\frac{\partial\Phi_{\xi\eta}}{\partial X^\lambda}
  \right]\,
  (\mathbf{i}_\xi\otimes\mathbf{i}_\eta\circ-\mathbf{i}_\alpha)\\
&= \mathbf{\Phi}\circ-\bnb -(\bnb\bdot\mathbf{n})\,\left(\mathbf{\Phi}\circ-\mathbf{n}\right)
   -(\mathbf{n}\cdot(\bnb\otimes\mathbf{\Phi}) )\circ-\mathbf{n}
\end{alignat*}

Thirdly, the full dimensional gradient is represented by the surface gradient
\begin{alignat*}{3}
     RHS
&   =\mathbf{\Phi}\circ-\left( \os{\Sigma}{\bnb}+\mbf{n}\frac{\pl}{\pl x^3} \right) -\left[ \left( \os{\Sigma}{\bnb}+\mbf{n}\frac{\pl}{\pl x^3} \right) \bdot\mathbf{n}\right]\,\left(\mathbf{\Phi}\circ-\mathbf{n}\right)\\
&   -\mathbf{n}\cdot\left[ \left( \os{\Sigma}{\bnb}+\mbf{n}\frac{\pl}{\pl x^3} \right)\otimes\mathbf{\Phi} \right]\circ-\mathbf{n}
    =\mbf{\Phi}\circ\os{\Sigma}{\bnb}+H\mbf{\Phi}\circ-\mbf{n}
\end{alignat*}
The proof is completed.

\subsubsection{Some integral identities for soft matter studies}

\cite{Yin-2008a} reported some kinds of novel integral identities that are taken as meaningful for soft matter studies. As a kind of applications, the intrinsic generalized Stokes formulas are utilized to deduce these identities as indicated in this subsection.

Firstly, the quantity termed as conjugate fundamental tensor by \cite{Yin-2008b} is introduced
\begin{equation*}
    |\mbf{K}|\mbf{K}^{-1}=:\Delta^i_j\mbf{g}_i\otimes\mbf{g}^j
\end{equation*}
where $\mbf{K}\triangleq b^i_j\mbf{g_i}\ots\mbf{g^j}$ is the curvature tensor, $[\Delta^i_j]$ denotes the adjugate matrix of $[b^i_j]$. Certainly, it should be pointed out that this quantity can only make sense in the case that $\mbf{K}$ is nonsingular, i.e. $\det[b^i_i]\neq 0$.

On $\Delta^i_j$, the following fundamental relations can be concluded
\begin{alignat*}{3}
& \Delta^i_j=b^s_s\delta^i_j-b^i_j,\quad &&b^i_j=\Delta^s_s\delta^i_j-\Delta^i_j\\
& e^{3ji}\Delta^l_i=-e^{3li}b^j_i,\quad  &&\eps^{3ji}\Delta^l_i=-\eps^{3li}b^j_i\\
& \nabla_i \Delta^i_j=0 &&
\end{alignat*}
The first two relationships can be directly verified. The last one is due to the Codazzi equation as indicated by \cite{Yin-2008b}. All of these relations play the essential role in the following deductions.

\begin{proposition}
\begin{alignat*}{3}
&   \oint_{\pl\Sigma}\,\mbf{\tau}\cdot\mbf{K}\circ-\mbf{\Phi}\,dl
    =\int_\Sigma\,\left( \mbf{n}\times \os{\Sigma}{\ol{\mbf{\nabla}}} \right)\circ-\mbf{\Phi}\,d\sigma
\end{alignat*}
where $\os{\Sigma}{\ol{\mbf{\nabla}}}:=\hat{L}^{ij}\mbf{g}_i\frac{\pl}{\pl x^j_\Sigma}$, $\hat{\mbf{L}}:=K_G\mbf{K}^{-1}$
\end{proposition}

\noindent\textbf{Proof}\quad

Firstly, it is worthy of mention that $K_G=\det[b^i_j]=:|\mbf{K}|$ and $\os{\Sigma}{\ol{\mbf{\nabla}}}=|\mbf{K}|\mbf{K}^{-1}\bcdot\os{\Sigma}{\bnabla}$.

As the application of the intrinsic generalized Stokes formula of the first kind, one has
\begin{equation*}
     \oint_{\pl\Sigma}\,\mbf{\tau}\cdot\mbf{K}\circ-\mbf{\Phi}\,dl
    =\int_\Sigma\,\left( \mbf{n}\times\os{\Sigma}{\mbf{\nabla}} \right)\bcdot(\mbf{K}\circ-\mbf{\Phi})\,d\sigma
\end{equation*}
with
\begin{alignat*}{3}
  \left( \mbf{n}\times\os{\Sigma}{\mbf{\nabla}} \right)\bcdot(\mbf{K}\circ-\mbf{\Phi})
&=\left( \mbf{n}\times\mbf{g}^l \right)\bcdot\left( \frac{\pl\mbf{K}}{\pl x^l_\Sigma}\circ-\mbf{\Phi}+\mbf{K}\circ-\frac{\pl\mbf{\Phi}}{\pl x^l_\Sigma}\right)=
\left( \mbf{n}\times\mbf{g}^l \right)\bcdot\left( \mbf{K}\circ-\frac{\pl\mbf{\Phi}}{\pl x^l_\Sigma}\right)\\
&=\eps^{3lk}b^j_k\mbf{g}_j\circ-\frac{\pl \mbf{\Phi}}{\pl x^l_\Sigma}
\end{alignat*}
thanks to
\begin{alignat*}{3}
  \left( \mbf{n}\times\mbf{g}^l \right)\bcdot \frac{\pl\mbf{K}}{\pl x^l_\Sigma}
&=\left( \mbf{n}\times\mbf{g}^l \right)\bcdot\left( \nabla_l b_{ij}\mbf{g}^i\otimes\mbf{g}^j
 +b_{ij}b^i_l\mbf{n}\otimes\mbf{g}^j+b_{ij}b^j_l\mbf{g}^i\otimes\mbf{n} \right) \\
&=\eps^{3li}\nabla_l b_{ij}\mbf{g}^j+\eps^{3li}(b_{ij}b^j_l)\mbf{n}=\mbf{0}
\end{alignat*}

On the other hand, one has
\begin{alignat*}{3}
  \left( \mbf{n}\times\os{\Sigma}{\ol{\bnabla}} \right)\circ-\mbf{\Phi}
&=\left[ \mbf{n}\times\left( |\mbf{K}|\mbf{K}^{-1}\bcdot\os{\Sigma}{\nabla} \right) \right]\circ-\mbf{\Phi}
 =\left[ \mbf{n}\times\left( \Delta^l_i\mbf{g}^i\frac{\pl}{\pl x^l_\Sigma}\right) \right]\circ-\mbf{\Phi}\\
&=\left[ \mbf{n}\times\left( \Delta^l_i\mbf{g}^i\right) \right]\circ-\frac{\pl\mbf{\Phi}}{\pl x^l_\Sigma}
 =\eps^{3ik}\Delta^l_i\mbf{g}_k\circ-\frac{\pl\mbf{\Phi}}{\pl x^l_\Sigma}
 =\eps^{3ij}\Delta^l_i\mbf{g}_j\circ-\frac{\pl\mbf{\Phi}}{\pl x^l_\Sigma}\\
&=-\eps^{3ji}\Delta^l_i\mbf{g}_j\circ-\frac{\pl\mbf{\Phi}}{\pl x^l_\Sigma}
 =\eps^{3li} b^j_i\mbf{g}_j\circ-\frac{\pl\mbf{\Phi}}{\pl x^l_\Sigma}
\end{alignat*}
It's the end of the proof.

\begin{proposition}
\begin{alignat*}{3}
&   \oint_{\pl\Sigma}\,(\mbf{\tau}\times\mbf{n})\cdot\hat{\mbf{L}}\circ-\mbf{\Phi}\,dl
    =\int_\Sigma\,\os{\Sigma}{\ol{\mbf{\nabla}}}\circ-\mbf{\Phi}\,d\sigma
   +\int_\Sigma\,2K_G(\mbf{n}\circ-\mbf{\Phi})\,d\sigma
\end{alignat*}
where $\hat{\mbf{L}}=|\mbf{K}|\mbf{K}^{-1}$.
\end{proposition}
\noindent\textbf{Proof}\quad

On the left hand side, one has
\begin{alignat*}{3}
    \oint_{\pl\Sigma}\,(\mbf{\tau}\times\mbf{n})\bcdot(|\mbf{K}|\mbf{K}^{-1})\circ-\mbf{\Phi}\,dl
&   =\oint_{\pl\Sigma}\,\left(\mbf{\tau}\times\mbf{n}\right)
     \bcdot\left(|\mbf{K}|\mbf{K}^{-1}\circ-\mbf{\Phi}\right)\,dl\\
&   =\int_\Sigma\,\left[ \os{\Sigma}{\bnabla}\bcdot (|\mbf{K}|\mbf{K}^{-1}\circ-\mbf{\Phi})+H\mbf{n}\bcdot
 (|\mbf{K}|\mbf{K}^{-1}\circ-\mbf{\Phi}) \right]\,d\sigma\\
&    =\int_\Sigma\, \os{\Sigma}{\bnabla}\bcdot (|\mbf{K}|\mbf{K}^{-1}\circ-\mbf{\Phi})\,d\sigma
\end{alignat*}

To deal with
\begin{alignat*}{3}
    \os{\Sigma}{\bnabla}\bcdot(|\mbf{K}|\mbf{K}^{-1}\circ-\mbf{\Phi})
&  =\mbf{g}^l\bcdot\frac{\pl}{\pl x^l_\Sigma}(|\mbf{K}|\mbf{K}^{-1}\circ-\mbf{\Phi})\\
&  =\mbf{g}^l\bcdot\left[ \frac{\pl}{\pl x^l_\Sigma}(|\mbf{K}|\mbf{K}^{-1})\circ-\mbf{\Phi}
+|\mbf{K}|\mbf{K}^{-1}\circ-\frac{\pl \mbf{\Phi} }{\pl x^l_\Sigma} \right]
\end{alignat*}
one deduces the second term on the right hand side as
\begin{alignat*}{3}
    \mbf{g}^l\bcdot\left( |\mbf{K}|\mbf{K}^{-1}\circ-\frac{\pl\mbf{\Phi}}{\pl x^l_\Sigma} \right)
&  =\left[ \mbf{g}^l\bcdot\left( |\mbf{K}|\mbf{K}^{-1} \right) \right] \circ-\frac{\pl\mbf{\Phi}}{\pl x^l_\Sigma}
   =\left( |\mbf{K}|\mbf{K}^{-1}\bcdot\mbf{g}^l \right) \circ-\frac{\pl\mbf{\Phi}}{\pl x^l_\Sigma}\\
&  =\left[ |\mbf{K}|\mbf{K}^{-1}\bcdot\left( \mbf{g}^l\frac{\pl}{\pl x^l_\Sigma} \right)  \right]\circ-\mbf{\Phi}
   =\os{\Sigma}{\ol{\bnabla}}\circ-\mbf{\Phi}
\end{alignat*}
and the first term on the right hand side as
\begin{alignat*}{3}
    \mbf{g}^l\bcdot\left[ \frac{\pl}{\pl x^l_\Sigma}(|\mbf{K}|\mbf{K}^{-1})\circ-\mbf{\Phi} \right]
&  =\mbf{g}^l\bcdot\left[ \frac{\pl}{\pl x^l_\Sigma}(\Delta^i_j\mbf{g}_i\otimes\mbf{g}^j)\circ-\mbf{\Phi} \right]\\
&  =\mbf{g}^l\bcdot\left[ \left( \nabla_l\Delta^i_j\mbf{g}_i\otimes\mbf{g}^j+\Delta^i_j b_{li}\mbf{n}\otimes\mbf{g}^j
+\Delta^i_j b^j_l\mbf{g}_i\otimes\mbf{n} \right)\circ-\mbf{\Phi} \right]\\
&  =( \nabla_l\Delta^l_j\mbf{g}^j + \Delta^i_j b^j_i\mbf{n})\circ-\mbf{\Phi}
   =(\delta^i_j b^s_s-b^i_j)b^j_i \mbf{n}\circ-\mbf{\Phi}
   =(b^j_jb^s_s-b^i_jb^j_i)\mbf{n}\circ-\mbf{\Phi}\\
&  =2K_G\mbf{n}\circ-\mbf{\Phi}
\end{alignat*}
The last identity is due to the relationship
\begin{equation*}
    K_G=\det[b^i_j]=\frac{1}{2}\delta^{ij}_{pq}b^p_ib^q_j
       =\frac{1}{2}\left|
       \begin{array}{cc}
       \delta^i_p & \delta^i_q\\
       \delta^j_p & \delta^j_q
       \end{array}
       \right|b^p_ib^q_j
       =\frac{1}{2}(b^j_jb^s_s-b^i_jb^j_i)
\end{equation*}
It's the end of the proof.

In studies by Yin with his collaborators (2005, 2008) on some integral identities, the following one plays the essential role
\begin{equation*}
    \int_{\Sigma}\,\os{\Sigma}{\bnb}\bdot(\mbf{\Theta}\circ-\mbf{\Phi}) d\sigma
   =\oint_{\pl\s}\,(\mbf{\tau}\ts\mbf{n})\bdot(\mbf{\Theta}\circ-\mbf{\Phi})\,dl,
   \quad\forall \mbf{\Theta}\in\mathscr{T}^r(\mbf{T\Sigma}),\,\forall \mbf{\Phi}\in\mathscr{T}^p(\mathbb{R}^3)
\end{equation*}
Its validity can be confirmed as soon as the intrinsic generalized Stokes formula of the second kind is adopted, namely
\begin{equation*}
    \oint_{\pl\s}\,(\mbf{\tau}\ts\mbf{n})\bdot(\mbf{\Theta}\circ-\mbf{\Phi})\,dl
   =\int_{\Sigma}\,\left[ \os{\Sigma}{\bnb}\bdot(\mbf{\Theta}\circ-\mbf{\Phi})
   +H\mbf{n}\bdot(\mbf{\Theta}\circ-\mbf{\Phi})  \right]\,d\sigma
   =\int_{\Sigma}\, \os{\Sigma}{\bnb}\bdot(\mbf{\Theta}\circ-\mbf{\Phi}) \,d\sigma
\end{equation*}

By other ways, one can do the following calculation, let $\mbf{\Theta}=\Theta^{ij}\mbf{g}_i\ots\mbf{g}_j$ without lost of the generality,
\begin{alignat*}{3}
     \int_{\Sigma}\,\os{\Sigma}{\bnb}\bdot(\mbf{\Theta}\circ-\mbf{\Phi})\, d\sigma
&   =\int_{\Sigma}\,\frac{1}{\sqrt{g}}\frac{\pl}{\pl x^s_\s}\left( \sqrt{g} \Theta^{sj}\mbf{g}_j\circ-\mbf{\Phi}
     \right)(x_\s,t)\,d\sigma\\
&   =\int_{\mathscr{D}_x}\, \frac{\pl}{\pl x^s_\s}\left( \sqrt{g} \Theta^{sj}\mbf{g}_j\circ-\mbf{\Phi}
     \right)(x_\s,t)\,d\sigma
    =\oint_{\pl\s}\,(\mbf{\tau}\ts\mbf{n})\bdot(\mbf{\Theta}\circ-\mbf{\Phi})\,dl
\end{alignat*}

The first identity is due to
\begin{alignat*}{3}
     \os{\Sigma}{\bnb}\bdot(\mbf{\Theta}\ots\mbf{\Phi})
&   =\mbf{g}^l\bdot\left[ \frac{\pl}{\pl x^l_\s}\left( \Theta^{ij}\mbf{g}_i\ots\mbf{g}_j \right)\circ-\mbf{\Phi}+\mbf{\Theta}\circ-\frac{\pl\mbf{\Phi}}{\pl
     x^l_\s}\right]\\
&   =\mbf{g}^l\bdot\left[\nb_l\Theta^{ij}\mbf{g}_i\ots\mbf{g}_j
+\Theta^{ij}(b_{li}\mbf{n}\ots\mbf{g}_j+b_{lj}\mbf{g}_i\ots\mbf{n}) \right]\circ-\mbf{\Phi}+\Theta^{ls}\mbf{g}_s\circ-\frac{\pl\mbf{\Phi}}{\pl
     x^l_\s}\\
&=\left(\nb_i\Theta^{ij}\mbf{g}_j+\Theta^{ij}b_{ij}\mbf{n}\right)\circ-\mbf{\Phi}
+\Theta^{ls}\mbf{g}_s\circ-\frac{\pl\mbf{\Phi}}{\pl
     x^l_\s}
\end{alignat*}
where
\begin{alignat*}{3}
    \nb_i\Theta^{ij}\mbf{g}_j+\Theta^{ij}b_{ij}\mbf{n}
&  =\left(\frac{\pl\Theta^{ij}}{\pl x^i_\s}+\Gamma^i_{is}\Theta^{sj}+\Gamma^j_{is}\Theta^{is} \right)\mbf{g}_j
   +\Theta^{ij}b_{ij}\mbf{n}\\
&  =\left( \frac{\pl\Theta^{sj}}{\pl x^s_\s}+\frac{1}{\sqrt{g}}\frac{\pl\sqrt{g}}{\pl x^s}\Theta^{sj}\right) \mbf{g}_j+\Theta^{sj}(\Gamma^k_{sj}\mbf{g}_k+b_{sj}\mbf{n})\\
&  =\frac{1}{\sqrt{g}}\frac{\pl}{\pl x^s_\s}(\sqrt{g}\Theta^{sj})\mbf{g}_j+\Theta^{sj}\frac{\pl\mbf{g}_j}{\pl x^s_\s}
   =\frac{1}{\sqrt{g}}\frac{\pl}{\pl x^s_\s}(\sqrt{g}\Theta^{sj}\mbf{g}_j)
\end{alignat*}
then
\begin{alignat*}{3}
    RHS=\frac{1}{\sqrt{g}}\left[ \frac{\pl}{\pl x^s_\s}(\sqrt{g}\Theta^{sj}\mbf{g}_j)\circ-\mbf{\Phi}
    +\sqrt{g}\theta^{sj}\mbf{g}_j\circ-\frac{\pl\mbf{\Phi}}{\pl x^s_\s} \right]
    =\frac{1}{\sqrt{g}}\frac{\pl}{\pl x^s_\s}(\sqrt{g}\Theta^{sj}\mbf{g}_j\circ-\mbf{\Phi})
\end{alignat*}

The last identity is essentially due to the Green formula. In detail, firstly one has the relation as the prototype
\begin{equation*}
    \int_{\mathscr{D}_x}\,\frac{\pl}{\pl x^s_\s}(\sqrt{g}a^s)\,d\sigma=\int_{\Sigma}\,(\mbf{\tau}\ts\mbf{n})\bdot\mbf{a}\,dl,
    \quad\forall\,\mbf{a}\in\mbf{T\Sigma}
\end{equation*}
As the left hand side is considered, it can be calculated as follows
\begin{alignat*}{3}
     \int_{\mathscr{D}_x}\,\frac{\pl}{\pl x^s_\s}(\sqrt{g}a^s)\,d\sigma
&    =\int_{\mathscr{D}_x}\,\left[ \frac{\pl}{\pl x^1_\s}(\sqrt{g}a^1)+\frac{\pl}{\pl x^2_\s}(\sqrt{g}a^2) \right] \,d\sigma
    =\int_{\pl\mathscr{D}_x}\left[ -\sqrt{g}a^2\mathbf{i}_1+\sqrt{g}a^1\mathbf{i}_2 \right]\bdot{\tau}\,dl\\
&   =\int^b_a\,\left[ -\sqrt{g}a^2 \dot{x}^1(t) + \sqrt{g}a^1 \dot{x}^2(t) \right]\,dt
    =\int^b_a\,\eps_{3ij}a^i\dot{x}^j(t)\,dt,\quad \eps_{3ij}=\sqrt{g}e_{3ij}\\
&   =\int^b_a\,\left[ \mbf{n},\, a^i\mbf{g}_i,\,\dot{x}^j(t)\mbf{g}_j \right]\,dt
    =\oint_{\pl\Sigma}\,\left[ \mbf{n},\, \mbf{a},\,\mbf{\tau} \right]\,dt
    =\oint_{\pl\Sigma}\,(\mbf{\tau}\ts\mbf{n})\bdot\mbf{a}\,dt
\end{alignat*}
Subsequently, the relation as the general type
\begin{equation*}
    \int_{\mathscr{D}_x}\, \frac{\pl}{\pl x^s_\s}\left( \sqrt{g} \Theta^{sj}\mbf{g}_j\circ-\mbf{\Phi}
     \right)(x_\s,t)\,d\sigma
    =\oint_{\pl\s}\,(\mbf{\tau}\ts\mbf{n})\bdot(\mbf{\Theta}\circ-\mbf{\Phi})\,dl
\end{equation*}
can be verified. The essential of the deduction is to transfer some indices of the tensor with respect to the local bases to the ones with respect to the canonical bases, namely
\begin{alignat*}{3}
    \sqrt{g} \Theta^{sj}\mbf{g}_j\circ-\mbf{\Phi}
   =\sqrt{g} \Theta^{sj}\mbf{g}_j\circ-
    \left( \Phi^{\beta}_{\cdot\gamma}\mathbf{g}_\beta\ots\mbf{g}^\gamma \right)
   =\sqrt{g} \Theta^{s\alpha}\mathbf{i}_\alpha\circ-\left( \Phi^{\lambda}_{\cdot\mu}\mathbf{i}_\lambda\ots\mathbf{i}^\mu \right)
\end{alignat*}
where let $\mbf{\Phi}\in\mathscr{T}^2(\mbf{T\Sigma})$ extent to $\mathscr{T}^2(\mathbb{R}^3)$ with the constrain $\mbf{\Phi}(\bdot,\mbf{n})=\mbf{\Phi}(\mbf{n},\bdot)=0\in\mathbb{R}$ then the transformation can be carried out
\begin{equation*}
    \Theta^{sj}\mbf{g}_j=\Theta^{s\beta}\mbf{g}_\beta:=\mbf{\Theta}(\mbf{g}^s,\mbf{g}^\beta)\mbf{g}_\beta
   =\mbf{\Theta}(\mbf{g}^s,(\mbf{g}^\beta,\mathbf{i}_\alpha)_{\mathbb{R}^3}\mathbf{i}^\alpha)\mbf{g}_\beta
   =\mbf{\Theta}(\mbf{g}^s,\mathbf{i}^\alpha)[(\mathbf{i}_\alpha,\mbf{g}^\beta)_{\mathbb{R}^3}\mbf{g}_\beta]
   =:\Theta^{s\alpha}\mathbf{i}_\alpha
\end{equation*}
Consequently, one can do the following deduction
\begin{alignat*}{3}
    \int_{\mathscr{D}_x}\, \frac{\pl}{\pl x^s_\s}\left( \sqrt{g} \Theta^{sj}\mbf{g}_j\circ-\mbf{\Phi}
    \right)\,d\sigma
&    =\left[ \int_{\mathscr{D}_x}\, \frac{\pl}{\pl x^s_\s}\left( \sqrt{g} \Theta^{s\alpha}
    \Phi^{\lambda}_{\cdot\mu}  \right)\,d\sigma\right]\,
    \mathbf{i}_\alpha\circ-\mathbf{i}_\lambda\ots\mathbf{i}^\mu\\
&   = \left[ \oint_{\pl\Sigma}\,(\mbf{n}\ts\mbf{\tau})\bdot\left(\Theta^{s\alpha}
    \Phi^{\lambda}_{\cdot\mu}\,\mbf{g}_s \right) \right]\,
    \mathbf{i}_\alpha\circ-\mathbf{i}_\lambda\ots\mathbf{i}^\mu\\
&   =\oint_{\pl\Sigma}\,(\mbf{n}\ts\mbf{\tau})\bdot\left[ \Theta^{s\alpha}
     \mbf{g}_s\ots\mathbf{i}_\alpha\circ-(\Phi^{\lambda}_{\cdot\mu}\mathbf{i}_\lambda\ots\mathbf{i}^\mu) \right]\,dl
\end{alignat*}
where the relation as the prototype is adopted.

The relations as the prototype and general type have been adopted directly by \cite{Yin-2005} and \cite{Yin-2008b}
respectively.

\subsubsection{A kind of ways to deduce governing equations for thin enough continuous mediums}

To study the representation of the natural law of momentum conservation for the continuous medium whose geometrical configuration can be taken as a surface, the so termed surface stress can be introduced
\begin{equation*}
    \mathbf{t}=t^i_{\cdot j}\boldsymbol{g}_i\otimes\boldsymbol{g}^j+t^i_{\cdot 3}\boldsymbol{g}_i\otimes\boldsymbol{n}
\end{equation*}

Subsequently,the momentum conservation can be set up in the integral form
\begin{equation*}
    \int_{\overset{t}{\Sigma}} \rho\,\boldsymbol{a}\,d\sigma
   =\oint_{\partial\overset{t}{\Sigma}} (\mbf{\tau}\times\mbf{n})\bdot\mathbf{t}\,dl
   +\int_{\overset{t}{\Sigma}} \boldsymbol{f}_\Sigma\,d\sigma
\end{equation*}
where $\mbf{f}_\Sigma$ denotes the distribution of the action imposed directly on the surface such as the weight, fraction and electromagnetic force. The differential equation of momentum conservation can be directly attained through the intrinsic generalized Stokes formula of the second kind
\begin{equation*}
   \rho\,\boldsymbol{a}=\overset{\Sigma}{\nabla}\bcdot\mathbf{t}+\boldsymbol{f}_\Sigma
\end{equation*}
with the component forms
\begin{alignat*}{3}
& \rho a_l
=\nb_s t^s_{\cdot l}-b_{sl}t^s_{\cdot 3}+f_l\\
& \rho a_n
=\nabla_st^s_{\cdot 3}+b^i_j t^j_{\cdot i}+f_n
\end{alignat*}

On the other hand, the moment of momentum conservation can be represented as
\begin{equation*}
    \int_{\overset{t}{\Sigma}} \rho\,\boldsymbol{a}\times\boldsymbol{\Sigma}\:\mathrm{d}\sigma
   =\oint_{\partial\overset{t}{\Sigma}} \left[(\mbf{\tau}\times\mbf{n})\bdot\mathbf{t}\right]\times\boldsymbol{\Sigma}\:\mathrm{d}l
   +\int_{\overset{t}{\Sigma}} \boldsymbol{f}_\Sigma\times\boldsymbol{\Sigma}\:\mathrm{d}\sigma
   +\int_{\overset{t}{\Sigma}} \boldsymbol{m}_\Sigma\:\mathrm{d}\sigma
\end{equation*}
with the differential form
\begin{equation*}
    \rho\,\boldsymbol{a}\times\boldsymbol{\Sigma}=\overset{\Sigma}{\bnabla}\bcdot(\mathbf{t}\times\boldsymbol{\Sigma})
    +\boldsymbol{f}_\Sigma\times\boldsymbol{\Sigma}+\boldsymbol{m}_\Sigma
    =\left[\, \left( \overset{\Sigma}{\bnabla}\bcdot\mathbf{t} \right)\times\boldsymbol{\Sigma}+\boldsymbol{g}^l\bcdot(\mathbf{t}\times\boldsymbol{g}_l)\,\right]
    +\boldsymbol{f}_\Sigma\times\boldsymbol{\Sigma}+\boldsymbol{m}_\Sigma
\end{equation*}
where $\mbf{m}_\Sigma$ denotes the surface force couple.

Substituting the governing equation of momentum conservation, one arrive at the governing equation of moment of momentum conservation
\begin{equation*}
   \mbf{0}
   =\boldsymbol{g}^l\bcdot(\mathbf{t}\times\boldsymbol{g}_l)+\mbf{m}_\Sigma
   =-t^{ij}\eps_{ij3}\boldsymbol{n}
   +\sqrt{g}(-t^2_{\cdot 3}\boldsymbol{g}^1+t^1_{\cdot 3}\boldsymbol{g}^2)+\mbf{m}_\Sigma,
   \quad g:=\det[g_{ij}]
\end{equation*}
Consequently, it can be concluded that the symmetry of the components of surface stress tensor on the tangent space, i.e. $t_{ij}=t_{ji}$, corresponds to the vanishing of the component of surface force couple in the surface normal direction. And the appearance of surface stress tensor in the surface normal direction, i.e. $t^i_{\cdot 3}\neq 0$, corresponds to the existence of components of surface force couple on the tangent space.

The governing equations of the statical force equilibrium of elastic plates and shells put forward by \cite{Chien-1941} are included in the above mentioned equation of momentum conservation. Comparatively, the deduction based on the intrinsic generalized Stokes formula of the second kind seems more compactly. Both \cite{Chien-1941} and \cite{Aris-1962} have introduced the concept of membrane or surface stress tensor in their studies on solids or fluids whose geometrical configurations can be taken as surfaces. Subsequently, the stress force can be represented as the surface divergence of the stress tensor and the differential equations of nature laws can be readily deduced from the integral representations through the intrinsic generalized Stokes formula of the second. However, as indicated by \cite{XXL-2013}, some kinds of actions on the boundary can not purely come down to the surface divergence, particularly as the continuous mediums have motions in the normal direction. In other words, the membrane or surface stress tensor may not be the universal representation of the force action on the boundary in all cases.

\subsubsection{A differential identity for vorticity dynamics}

\begin{proposition}
On any deformable smooth surface, the following identity is keeping valid
\begin{equation*}
    \mbf{n}\bcdot\left[ \os{\Sigma}{\bnabla}\times(\mbf{n}\times\mbf{\Phi}) \right]
   =\left( \mbf{n}\times \os{\Sigma}{\bnabla}\right)\bcdot(\mbf{n}\times\mbf{\Phi})
   =\os{\Sigma}{\bnabla}\bcdot\mbf{\Phi}+H\mbf{n}\bcdot\mbf{\Phi},\quad
   \forall\,\mbf{\Phi}\in\mathscr{T}^p(\mathbb{R}^3)
\end{equation*}
\end{proposition}
\noindent\textbf{Proof}\quad

This identity can be readily proved through the utilizations of the intrinsic generalized Stokes formula of the first kind
\begin{equation*}
    \oint_{\pl\Sigma}\,\mbf{\tau}\bdot(\mbf{n}\times\mbf{\Phi})\,dl
   =\int_\Sigma\,\left( \mbf{n}\times \os{\Sigma}{\bnabla}\right)\bcdot(\mbf{n}\times\mbf{\Phi})\,d\sigma
\end{equation*}
and the one of the second kind
\begin{equation*}
    \oint_{\pl\Sigma}\,(\mbf{\tau}\times\mbf{n})\bdot\mbf{\Phi}\,dl
   =\int_\Sigma\,\left( \os{\Sigma}{\bnabla}\bcdot\mbf{\Phi}+H\mbf{n}\bcdot\mbf{\Phi} \right)\,d\sigma
\end{equation*}
accompanying with the fundamental relationship
\begin{equation*}
    \mbf{\tau}\bdot(\mbf{n}\times\mbf{\Phi})=(\mbf{\tau}\times\mbf{n})\bdot\mbf{\Phi},\quad
    \forall\,\mbf{\Phi}\in\mathscr{T}^p(\mathbb{R}^3)
\end{equation*}
It is the end of the proof. On the other hand, one can prove this identity through direct calculations.

In fluid mechanics, the nature law of momentum conservation is represented by so called Navier-Stokes equation (NSE)
\begin{equation*}
    \rho\mbf{a}=\bnb\Pi-\mu\bnb\times\mbf{\omega}+\rho\mbf{f}_m
\end{equation*}
where $\Pi:=-p+(\lambda+2\mu)\theta$ is the dilation quantity, $\lambda$ and $\mu$ denote different viscous coefficients, $\mbf{\omega}:=\bnb\times\mbf{V}$ is the vorticity. To deduce the representation of the flux of the dilation quantity on any smooth deformable boundary, $\bdot\mbf{n}$ is taken on both side of NSE
\begin{alignat*}{3}
    \frac{\pl\Pi}{\pl\mbf{n}}:=\mbf{n}\bdot\bnb\Pi
   &=\rho\mbf{n}\bdot\mbf{a}+\mu\mbf{n}\bdot(\bnb\times\mbf{\omega})-\rho\mbf{n}\bdot\mbf{f}_m\\
   &=\rho\mbf{n}\bdot\mbf{a}+\mu(\mbf{n}\times\bnb)\bdot\mbf{\omega}-\rho\mbf{n}\bdot\mbf{f}_m
    =\rho\mbf{n}\bdot\mbf{a}+\mu\left( \mbf{n}\times\os{\Sigma}{\bnb} \right)\bdot\mbf{\omega}-\rho\mbf{n}\bdot\mbf{f}_m
\end{alignat*}
On the other hand, $\times\mbf{n}$ is taken on both side of NSE to deduce the representation of the flux of the vorticity on the boundary
\begin{alignat*}{3}
    \mu\frac{\pl\mbf{\omega}}{\pl\mbf{n}}
&   =\rho\mbf{n}\times\mbf{a}-\mbf{n}\times(\bnb\Pi)
    +\mu(\bnb\otimes\mbf{\omega})\bdot\mbf{n}+\rho\mbf{f}_m\times\mbf{n}\\
&   =\rho\mbf{n}\times\mbf{a}-\left( \mbf{n}\times\overset{\Sigma}{\bnb} \right)\Pi
    +\mu(\bnb\otimes\mbf{\omega})\bdot\mbf{n}+\rho\mbf{f}_m\times\mbf{n}
\end{alignat*}
where the following identity is adopted
\begin{equation*}
    (\bnb\times\mbf{\omega})\times\mbf{n}=\mbf{n}\bdot(\bnb\otimes\mbf{\omega})
    -(\bnb\otimes\mbf{\omega})\bdot\mbf{n}
\end{equation*}

Furthermore, one has the following relationships
\begin{alignat*}{3}
    (\bnb\otimes\mbf{\omega})\bdot\mbf{n}
&   =\os{\Sigma}{\bnb}(\mbf{\omega}\bdot\mbf{n})-\left[ \mbf{\omega}\bdot\left( \os{\Sigma}{\bnb}\otimes\mbf{n} \right)
    +\left( \os{\Sigma}{\bnb}\bdot\mbf{\omega} \right)\mbf{n} \right]
    =\os{\Sigma}{\bnb}(\mbf{\omega}\bdot\mbf{n})-\os{\Sigma}{\bnb}\bdot(\mbf{\omega}\otimes\mbf{n})
\end{alignat*}
thanks to $\bnb\bdot\mbf{\omega}=0$ and
\begin{equation*}
    \mbf{\xi}\bdot\left( \os{\Sigma}{\bnb}\otimes\mbf{\eta}\right)
   +\left( \os{\Sigma}{\bnb}\bdot\mbf{\xi}\right)\mbf{\eta}
   =\os{\Sigma}{\bnb}\bdot(\mbf{\xi}\otimes\mbf{\eta}),\quad
   \forall\,\mbf{\xi},\,\mbf{\eta}\in\mathbb{R}^3
\end{equation*}

Finally, one arrives at the representation
\begin{alignat*}{3}
    \mu\frac{\pl\mbf{\omega}}{\pl\mbf{n}}
&   =\rho\mbf{n}\times\mbf{a}-\left( \mbf{n}\times\overset{\Sigma}{\bnb} \right)\Pi
    +\mu(\bnb\otimes\mbf{\omega})\bdot\mbf{n}+\rho\mbf{f}_m\times\mbf{n}\\
&   =\rho\mbf{n}\times\mbf{a}-\left( \mbf{n}\times\overset{\Sigma}{\bnb} \right)\Pi
    +\left( \mbf{n}\times\overset{\Sigma}{\bnb} \right)\bdot\left[ (\mbf{\omega}\times\mbf{n})\otimes\mbf{n} \right]
    +\mu(\bnb\otimes\mbf{\omega})\bdot\mbf{n}+\rho\mbf{f}_m\times\mbf{n}\\
&   +\mu\left[ \os{\Sigma}{\bnb}(\mbf{\omega}\bdot\mbf{n})+H(\mbf{\omega}\bdot\mbf{n})\mbf{n} \right]
\end{alignat*}
where $\mu\left[ \os{\Sigma}{\bnb}(\mbf{\omega}\bdot\mbf{n})+H(\mbf{\omega}\bdot\mbf{n})\mbf{n} \right]$ is the additional term contributed purely by the deformation of the boundary and equals to zero for any flow on the plane. Its mechanical meaning can be revealed by the following integral relation
\begin{equation*}
    \int_\Sigma\,\mu\frac{\pl\mbf{\omega}}{\pl\mbf{n}}\,d\sigma\sim
    \int_\Sigma\,\mu\left[ \os{\Sigma}{\bnb}(\mbf{\omega}\bdot\mbf{n})+H(\mbf{\omega}\bdot\mbf{n})\mbf{n} \right]\,d\sigma=\oint_{\pl\Sigma}\,\mu(\mbf{\tau}\times\mbf{n})(\mbf{\omega}\bdot\mbf{n})\,dl
\end{equation*}

The representation of the vorticity flux on the fixed solid boundary can be referred to the monograph of \cite{WJZ-2005M}.

\subsection{Primary properties of deformation gradient tensor for thin enough continuous mediums}

The deformation gradient tensor plays the essential role in the whole theory of continuous mediums without regard to those geometrical configurations are bulks corresponding to Euclid manifolds or surfaces to Riemann manifolds. The primary properties of the deformation gradient tensor for continuous mediums whose geometrical configurations (termed briefly as \emph{the surface deformation theory} hereinafter) are surfaces can be concluded as follows.
\begin{proposition}[properties of deformation gradient for surface deformation theory]
\begin{alignat*}{3}
& \dot{\mbf{F}}=\mbf{L}\bdot\mbf{F},\quad && \mbf{L}:=\mbf{V}\otimes\os{\Sigma}{\bnb}\\
& \dot{\overline{|\mbf{F}|}}=\theta |\mbf{F}|, \quad && \theta: =\mbf{V}\bdot\os{\Sigma}{\bnb}
\end{alignat*}
\end{proposition}
where the deformation gradient with its determinant are defined as
\begin{equation*}
    \mbf{F}:=\frac{\pl x_\Sigma^i}{\pl \xi^A_\Sigma}(\xi,t)\mbf{g}_i\otimes\mbf{G}^A,\quad
    |\mbf{F}|=\frac{\sqrt{g}}{\sqrt{G}}\det\left[ \frac{\pl x_\Sigma^i}{\pl\xi^A}(\xi_\Sigma,t)\right]
\end{equation*}

Firstly, we put forward the following lemma for $g:=\det[g_{ij}]$
\begin{lemma}[Some identities on the determinant of the metric tensor]
\begin{equation*}
    \frac{1}{g}\frac{\pl g}{\pl x^l_\Sigma}(x_\Sigma,t)=g^{ij}\frac{\pl g_{ij}}{\pl x^l_\Sigma}(x_\Sigma,t);\quad
    \frac{1}{g}\frac{\pl g}{\pl t}(x_\Sigma,t)=g^{ij}\frac{\pl g_{ij}}{\pl t}(x_\Sigma,t)
\end{equation*}
\end{lemma}
\noindent\textbf{Proof of the lemma}\quad
To consider
\begin{equation*}
    g:=\det[g_{ij}]=\sum_{s=1}^m\,\Delta^{is}g_{is},\quad \forall\,i=1,\cdots,m
\end{equation*}
where $\Delta^{is}$ denotes the element in the $i-th$ row and $s-th$ column of the conjugate matrix of $[g_{pq}]$.
Then, the determinant can be represented as
\begin{equation*}
    g=g(\{g_{ij}\,\mbox{included}\})\quad
    \mbox{s.t.}\quad
    \frac{\pl g}{\pl g_{ij}}(\{g_{ij}\,\mbox{included}\})=\Delta^{ij}\neq 0
\end{equation*}
It means that if $g_{ij}$ is included in the representation of $g$ then $\Delta^{ij}\neq 0$ and if $g$ does not include $g_{ij}$ then $\Delta^{ij}=0$  .

Subsequently, one can do the deduction
\begin{alignat*}{3}
    \frac{\pl g}{\pl x^l_\Sigma}=\sum_{ g_{ij}\,\mbox{included} }\,\frac{\pl g}{\pl g_{ij}}
    \frac{\pl g_{ij}}{\pl x^l_\Sigma}(x_\Sigma,t)
    =\sum_{ g_{ij}\,\mbox{included} }\,\Delta^{ij} \frac{\pl g_{ij}}{\pl x^l_\Sigma}(x_\Sigma,t)
&   =\sum_{p,q=1}^m\,\Delta^{pq} \frac{\pl g_{pq}}{\pl x^l_\Sigma}(x_\Sigma,t)\\
&   =gg^{pq} \frac{\pl g_{pq}}{\pl x^l_\Sigma}(x_\Sigma,t)
\end{alignat*}
The second identity can be proved similarly. As a corollary, one has the identity
\begin{equation*}
    \Gamma^i_{ij}\triangleq g^{ik}\Gamma_{ij,k}=\frac{1}{\sqrt{g}}\frac{\pl\sqrt{g}}{\pl x^j_\s}(x_\s,t),
    \quad \Gamma_{ij,k}\triangleq \frac{1}{2}\left( \frac{\pl g_{ik}}{\pl x^j_\s}+
     \frac{\pl g_{jk}}{\pl x^i_\s}-  \frac{\pl g_{ij}}{\pl x^k_\s} \right)(x_\s,t)
\end{equation*}

\noindent\textbf{Proof of the proposition}\quad
To calculate the material derivative of the deformation gradient tensor
\begin{alignat*}{1}
\dot{\mbf{F}}&=\dot{\overline{\frac{\pl x^i}{\pl \xi^A}(\xi,t)\mbf{g}_i(x,t)\otimes \mbf{G}^A(\xi)}}
=\dot{\overline{\frac{\pl x^i}{\pl \xi^A}(\xi,t)}}\mbf{g}_i\otimes G^A+\frac{\pl x^i}{\pl \xi^A}(\xi,t)\dot{\overline{\mbf{g}_i(x,t)}}\otimes \mbf{G}^A
\end{alignat*}
where
\begin{alignat*}{1}
\dot{\overline{\frac{\pl x^i}{\pl \xi^A}(\xi,t)}} &= \frac{\pl x^i}{\pl \xi^A\pl t}(\xi,t)=:\frac{\pl \dot{x}^i}{\pl\xi^A}(\xi,t)=\frac{\pl x^s}{\pl\xi^A}(\xi,t)\frac{\pl\dot{x}^i}{\pl x^s}(x,t)\\
\dot{\overline{\mbf{g}_i(x,t)}}&=\frac{\pl \mbf{g}_i}{\pl t}(x,t)+\dot{x^s_\s}\frac{\pl \mbf{g}_i}{\pl x^s_\s}(x,t)=\frac{\pl}{\pl x^i_\s}\left(\frac{\pl \mbf{\Sigma} }{\pl t}\right)(x,t)+\dot{x^s_\s}\frac{\pl \mbf{g}_s}{\pl x^i_\s}(x,t)
\end{alignat*}
Subsequently, one has
\begin{alignat*}{1}
\dot{\mbf{F}}&=\frac{\pl x_\s^s}{\pl \xi^A}\left[\frac{\pl\dot{x}_\s^i}{\pl x_\s^s}(x_\s,t)\mbf{g}_i\otimes \mbf{G}^A
    +\frac{\pl }{\pl x_\s^s}\left(\frac{\pl \mbf{\Sigma}}{\pl t}\right)(x_\s,t)\otimes \mbf{G}^A
    +\dot{x_\s}^i\frac{\pl\mbf{g}_i}{\pl x_\s^s}(x_\s,t)\otimes \mbf{G}^A\right]\\
& =\left[\frac{\pl}{\pl x_\s^s}\left(\frac{\pl\mbf{\Sigma}}{\pl t}\right)(x_\s,t)
    +\frac{\pl\dot{x}_\s^i}{\pl x_\s^s}(x_\s ,t)\mbf{g}_i+\dot{x}_\s^i\frac{\pl \mbf{g}_i}{\pl x_\s^s}(x_\s,t)\right]
    \otimes\left[\frac{\pl x_\s^s}{\pl\xi^A}(\xi,t)\mbf{G}^A\right]\\
& =\left[\frac{\pl}{\pl x_\s^s}\left(\frac{\pl \mbf{\s}}{\pl t}+\dot{x}_\s^i\mbf{g}_i\right)(x_\s,t)\otimes \mbf{g^s}\right]
    \mbf{\cdot}\left[\frac{\pl x_\s^t}{\pl \xi^A}(\xi,t)\mbf{g}_t\otimes \mbf{G}^A\right]\\
& =\left(\mbf{V}\otimes\os{\Sigma}{\bnb}\right)\mbf{\cdot}\mbf{F}
\end{alignat*}

To calculate the material derivative of the determinant of the deformation gradient tensor
\begin{alignat*}{1}
\dot{\overline{|\mbf{F}|}}&=\dot{\overline{\frac{\sqrt{g}(x_\s,t)}{\sqrt{G}(\xi,t)i}
    \det\left[\frac{\pl x_\s^i}{\pl\xi_\s^A}\right](\xi_\s,t)}}\\
& =\frac{1}{\sqrt{G}}\dot{\overline{\sqrt{g}(x_\s,t)}}\det\left[\frac{\pl x_\s^i}{\pl\xi_\s^A}\right](\xi_\s,t)
    +\frac{\sqrt{g}}{\sqrt{G}}\dot{\overline{\det\left[\frac{\pl x_\s^i}{\pl\xi_\s^A}\right](\xi_\s,t)}}
\end{alignat*}
where
\begin{alignat*}{1}
\dot{\overline{\sqrt{g}(x_\s,t)}}&=\frac{\pl\sqrt{g}}{\pl t}(x_\s,t)+\dot{x_\s^s}\frac{\pl \sqrt{g}}{\pl \dot{x}_\s^s}(x_\s,t)\\
&=\sqrt{g}\left[\frac{1}{\sqrt{g}}\frac{\pl\sqrt{g}}{\pl t}(x_\s,t)+\dot{x}_\s^s\frac{1}{\sqrt{g}}\frac{\pl \sqrt{g}}{\pl x_\s^s}(x_\s,t)\right]\\
&=\sqrt{g}\left[\frac{1}{\sqrt{g}}\frac{\pl\sqrt{g}}{\pl t}(x_\s,t)+\Gamma_{st}^t\dot{x}_\s^s\right]
\end{alignat*}
and
\begin{equation*}
\dot{\overline{\det\left[\frac{\pl x_\s^i}{\pl \xi_\s^A}\right](\xi_\s,t)}} =\frac{\pl \dot{x}_\s^s}{\pl x_\s^s}(x_\s,t)\det\left[\frac{\pl x_\s^i}{\pl \xi_\s^A}\right](\xi_\s,t)
\end{equation*}
Subsequently, one has
\begin{alignat*}{1}
\dot{\overline{|\mbf{F}|}}=|\mbf{F}|\cdot\left[\frac{1}{\sqrt{g}}\frac{\pl\sqrt{g}}{\pl t}(x_\s,t)+\frac{\pl \dot {x}_\s^s}{\pl x_\s^s}(x_\s,t)+\Gamma_{st}^s\dot{x}_\s^t\right]
=|\mbf{F}|\cdot\left[\frac{1}{\sqrt{g}}\frac{\pl\sqrt{g}}{\pl t}(x_\s,t)+\nabla_s\dot{x}_\s^s\right]
\end{alignat*}

On the other hand, the divergence of the velocity can be calculated as follows
\begin{alignat*}{1}
\mbf{V}\bdot\os{\Sigma}{\bnb}&=\frac{\pl \mbf{V}}{\pl x_\s^l}(x_\s,t)\mbf{\cdot} \mbf{g}^l
    =\frac{\pl}{\pl x^l_\s}\left(\frac{\pl \mbf{\s}}{\pl t}+\dot{x}_\s^s \mbf{g}_s\right)(x_\s,t)\mbf{\cdot}\mbf{g}^l\\
& =\mbf{g}^l\mbf{\cdot}\frac{\pl \mbf{g}_l}{\pl t}(x_\s,t)+\nabla_s\dot{x}_\s^s
  =g^{lk}\mbf{g}_k\mbf{\cdot}\frac{\pl \mbf{g}_l}{\pl t}(x_\s,t)+\nabla_s\dot{x}_\s^s\\
& =\frac{1}{2}g^{lk}\frac{\pl g_{lk}}{\pl t}(x_\s,t)+\nabla_s\dot{x}_\s^s
  =\frac{1}{2}\frac{1}{g}\frac{\pl g}{\pl t}(x_\s,t)+\nabla_s\dot{x}_\s^s
  =\frac{1}{\sqrt{g}}\frac{\pl\sqrt{g}}{\pl t}(x_\s,t)+\nabla_s\dot{x}_\s^s
\end{alignat*}
This ends the proof.

In addition, we give a proof of the identity that has been adopted previously
\begin{lemma}
\begin{equation*}
\dot{\overline{\det\left[\frac{\pl x_\s^i}{\pl \xi_\s^A}\right](\xi_\s,t)}} =\frac{\pl \dot{x}_\s^s}{\pl x_\s^s}(x_\s,t)\det\left[\frac{\pl x_\s^i}{\pl \xi_\s^A}\right](\xi_\s,t)
\end{equation*}
\end{lemma}
\noindent\textbf{Proof}\quad The determinant of an matrix can be represented through the permutation operator
\begin{alignat*}{3}
    \det\left[\frac{\pl x^i_\s}{\pl \xi^A_\s}\right](\xi_\s,t)
&  =\sum_{\sigma\in P_m}\,sgn\sigma\,\left[
    \frac{\pl x^1_\s}{\pl \xi^{\sigma(1)}_\s}\cdots \frac{\pl x^m_\s}{\pl \xi^{\sigma(m)}_\s}
    \right](\xi_\s,t)
\end{alignat*}
Subsequently, one has
\begin{alignat*}{3}
&   \dot{\overline{\det\left[\frac{\pl x_\s^i}{\pl \xi_\s^A}\right](\xi_\s,t)}}\\
&   =\sum_{\sigma\in P_m}\,sgn\sigma\,\left[
    \os{\bdot}{\ol{\frac{\pl x^1_\s}{\pl \xi^{\sigma(1)}_\s}}} \frac{\pl x^2_\s}{\pl \xi^{\sigma(2)}_\s} \cdots \frac{\pl x^m_\s}{\pl \xi^{\sigma(m)}_\s} +\cdots+
    \frac{\pl x^1_\s}{\pl \xi^{\sigma(1)}_\s} \cdots \frac{\pl x^{m-1}_\s}{\pl \xi^{\sigma(m-1)}_\s}
    \os{\bdot}{\ol{\frac{\pl x^m_\s}{\pl \xi^{\sigma(m)}_\s}}}
    \right](\xi_\s,t)
\end{alignat*}
To deal with the first term on the right hand side
\begin{alignat*}{3}
&   \sum_{\sigma\in P_m}\,sgn\sigma\,\left[
    \os{\bdot}{\ol{\frac{\pl x^1_\s}{\pl \xi^{\sigma(1)}_\s}}} \frac{\pl x^2_\s}{\pl \xi^{\sigma(2)}_\s} \cdots \frac{\pl x^m_\s}{\pl \xi^{\sigma(m)}_\s} \right](\xi_\s,t)
   =\sum_{\sigma\in P_m}\,sgn\sigma\,\left[
    \frac{\pl \dot{x}^1_\s}{\pl \xi^{\sigma(1)}_\s} \frac{\pl x^2_\s}{\pl \xi^{\sigma(2)}_\s} \cdots \frac{\pl x^m_\s}{\pl \xi^{\sigma(m)}_\s} \right](\xi_\s,t)\\
&  =\sum_{\sigma\in P_m}\,sgn\sigma\,\left[
    \left(\frac{\pl \dot{x}^1_\s}{\pl x^s_\s}(x_\s,t)\frac{\pl x^s_\s}{\pl \xi^{\sigma(1)}_\s}(\xi_\s,t)\right) \frac{\pl x^2_\s}{\pl \xi^{\sigma(2)}_\s}(\xi_\s,t) \cdots \frac{\pl x^m_\s}{\pl \xi^{\sigma(m)}_\s}(\xi_\s,t) \right]\\
&  =\frac{\pl \dot{x}^1_\s}{\pl x^s_\s}(x_\s,t) \sum_{\sigma\in P_m}\,sgn\sigma\,\left[
    \frac{\pl x^s_\s}{\pl \xi^{\sigma(1)}_\s} \frac{\pl x^2_\s}{\pl \xi^{\sigma(2)}_\s} \cdots \frac{\pl x^m_\s}{\pl \xi^{\sigma(m)}_\s} \right](\xi_\s,t)
    =\frac{\pl \dot{x}^1_\s}{\pl x^1_\s}(x_\s,t) \det\left[\frac{\pl x_\s^i}{\pl \xi_\s^A}\right](\xi_\s,t)
\end{alignat*}
It is evident that the other terms can be similarly processed. Then the proof is completed.

It should be pointed out that the concluded properties of the deformation gradient tensor with the related lemmas in this subsection are keeping valid for continuous mediums whose geometrical configurations are surfaces with arbitrary finite dimensions. Therefore, the related results can be taken as the extension of the surface deformation theory with respect to the dimensionality of the surface is two as put forward by \cite{XXL-2013}.

In the monograph \emph{Vectors, Tensors and the Basic Equations of Fluid Mechanics}, \cite{Aris-1962} expatiated on the equations governing two dimensional flows on an arbitrary fixed surface. It is worthy of mention that in Aris' monograph, the following relation was adopted
\begin{equation*}
\dot{\overline{\det\left[\frac{\pl x_\s^i}{\pl \xi_\s^A}\right](\xi_\s,t)}}
=
\left[ \frac{\pl \dot{x}_\s^s}{\pl x_\s^s}(x_\s,t)+\Gamma^s_{sj}\dot{x}^j_\s \right]
\det\left[\frac{\pl x_\s^i}{\pl \xi_\s^A}\right](\xi_\s,t)
=(\nb_s\dot{x}^s_\s) \det\left[\frac{\pl x_\s^i}{\pl \xi_\s^A}\right](\xi_\s,t)
\end{equation*}
However, it is not true. Consequently, Aris attained transport equation on the fixed surface takes the following form, see $(10.12.9)$ in the monograph,
\begin{equation*}
    \frac{d}{dt}\int_{\os{t}{\Sigma}}\,\Phi d\sigma
   =\int_{\os{t}{\Sigma}}\,\left[ \dot{\Phi}+\Phi\left( \nb_s V^s+\frac{\dot{g}}{2g}\right) \right]d\sigma,
   \quad V^s:=\dot{x}^s_\s
\end{equation*}
In fact, the true one should be
\begin{equation*}
    \frac{d}{dt}\int_{\os{t}{\Sigma}}\,\Phi d\sigma
   =\int_{\os{t}{\Sigma}}\,\left[ \dot{\Phi}+\Phi\left( \frac{\pl V^s}{\pl x^s_\s}(x_\s,t)+\frac{\dot{g}}{2g}\right) \right]d\sigma
   =\int_{\os{t}{\Sigma}}\,\left[ \dot{\Phi}+\Phi \nb_sV^s \right]d\sigma
\end{equation*}
accompanying with with the relation
\begin{equation*}
    \frac{\dot{g}}{2g}=\frac{1}{2g}\frac{\pl g}{\pl x^i_\s}(x_\s,t)\dot{x}^i
    =\frac{1}{2g}\frac{\pl g}{\pl x^i_\s}(x_\s,t)V^i=\Gamma^s_{si}V^i
\end{equation*}

The case studied by \cite{Aris-1962} is the two dimensional flow on an arbitrary fixed surface. The surface deformation theory put forward by \cite{XXL-2013} is available to two dimensional flows either on fixed surfaces or on deformable ones. Furthermore, a theoretical framework of vorticity dynamics for two dimensional flows on fixed surfaces has been provided recently by \cite{XXL-2013-arXiv}.

\subsection{Strain tensor on an arbitrary deformable surface}

For any motion/deformation of continuous mediums, the strain tensor on an arbitrary deformable smooth surface/boundary can take the following representation
\begin{proposition}[Stain tensor on an arbitrary deformable surface]
\begin{alignat*}{3}
    \mbf{D}
&  \triangleq\frac{1}{2}(\mbf{V}\otimes\bnb+\bnb\otimes\mbf{V})\\
&  =\left( \theta-\os{\Sigma}{\bnb}\bdot\mbf{V} \right)\mbf{n}\otimes\mbf{n}
   +\frac{1}{2}[(\mbf{\omega}+\mbf{W})\times\mbf{n}]\otimes\mbf{n}
   +\frac{1}{2}\mbf{n}\otimes[(\mbf{\omega}+\mbf{W})\times\mbf{n}]+\os{\Sigma}{\mbf{D}}
\end{alignat*}
where $\os{\Sigma}{\mbf{D}}\triangleq\left( \mbf{V}\otimes\os{\Sigma}{\bnb}+\os{\Sigma}{\bnb}\otimes\mbf{V} \right)/2$
is the strain of the boundary, $\mbf{W}:=-\left( \os{\Sigma}{\nb}V^3+\mbf{V}\bdot\mbf{K}\right)\times\mbf{n}$ is
purely determined by the boundary.
\end{proposition}

\noindent\textbf{Proof}\quad
Firstly, we introduce the intrinsic decomposition of any tensor field with respect to any direction
\begin{lemma}[Intrinsic decomposition with respect to any direction]
\begin{equation*}
\mbf{\Phi}=\left\{
\begin{array}{l}
\mbf{e}\ots(\mbf{e},\mbf{\Phi})_{\mathbb{R}^3}-[\mbf{e},[\mbf{e},\mbf{\Phi}]]\\
(\mbf{\Phi},\mbf{e})_{\mathbb{R}^3}\ots\mbf{e}-[[\mbf{\Phi},\mbf{e}],\mbf{e}]
\end{array}\quad
\forall\,|\mbf{e}|_{\mathbb{R}^3}=1,\quad\, \forall\mbf{\Phi}\in\mathscr{T}^p(\mathbb{R}^3)
\right.
\end{equation*}
\end{lemma}
It can be directly verified.

As an application, one has
\begin{alignat*}{1}
\mbf{V}\otimes \bnabla&=(\mbf{V}\otimes \bnabla,\mbf{n})\otimes \mbf{n}-\left[[\mbf{V}\otimes \bnabla,\mbf{n}],\mbf{n}\right]
\end{alignat*}
with the relations
\begin{alignat*}{1}
(\mbf{V}\otimes\bnabla,\mbf{n})_{\mathbb{R}^3} &=(\mbf{V}\otimes\bnabla-\bnabla\otimes\mbf{V},\mbf{n})_{\mathbb{R}^3}
    +(\bnabla\otimes\mbf{V},\mbf{n})_{\mathbb{R}^3}
    =\mbf{\omega}\times\mbf{n}+(\bnabla\otimes\mbf{V},\mbf{n})_{\mathbb{R}^3}\\
(\bnabla\otimes\mbf{V},\mbf{n})_{\mathbb{R}^3}&=\left(\overset{\s}{\bnabla}\otimes\mbf{V}+
    \mbf{n}\otimes\frac{\pl \mbf{V}}{\pl x^3}(x_\s,t),\mbf{n}\right)_{\mathbb{R}^3}
    =\left(\overset{\s}{\bnabla}\otimes\mbf{V},\mbf{n}\right)_{\mathbb{R}^3}+\frac{\pl V^3}{\pl x^3}(x_\s,t)\mbf{n}
\end{alignat*}
In addition, one has
\begin{alignat*}{1}
(\bnabla\otimes\mbf{V},\mbf{n})_{\mathbb{R}^3}
    &=\left(\os{\s}{\bnabla}\otimes \mbf{V},\mbf{n}\right)_{\mathbb{R}^3}
     +\left(\theta-\os{\s}{\bnabla}\bcdot\mbf{V}\right)\mbf{n}
\end{alignat*}
due to
\begin{alignat*}{1}
\theta :=\bnabla\bcdot\mbf{V}=\os{\s}{\bnabla}\bcdot\mbf{V}+\mbf{n}\bcdot\frac{\pl \mbf{V}}{\pl x^3}(x,t)
=\os{\s}{\bnabla}\bcdot\mbf{V}+\frac{\pl V^3}{\pl x^3}
\end{alignat*}
Furthermore, one has based on the intrinsic decomposition
\begin{alignat*}{1}
\left(\os{\s}{\bnabla}\otimes \mbf{V},\mbf{n}\right)_{\mathbb{R}^3}
    &=\mbf{n}\otimes\left(\mbf{n},\left(\os{\s}{\bnabla}\otimes\mbf{V},\mbf{n}\right)_{\mathbb{R}^3}\right)_{\mathbb{R}^3}
     -\left[\mbf{n},\left[\mbf{n},\left(\os{\s}{\bnabla}\otimes\mbf{V},\mbf{n}\right)_{\mathbb{R}^3}\right]\right]\\
    &=-\left[\mbf{n},\left[\mbf{n},\left(\os{\s}{\bnabla}\otimes\mbf{V},\mbf{n}\right)_{\mathbb{R}^3}\right]\right]
    =:\mbf{W}\times\mbf{n}
\end{alignat*}
where
\begin{alignat*}{1}
\mbf{W}:=-\left(\os{\s}{\bnabla}\otimes\mbf{V},\mbf{n}\right)_{\mathbb{R}^3}\times\mbf{n}
    =-\left(\os{\s}{\bnabla}V^3+\mbf{V}\bcdot\mbf{K}\right)\times\mbf{n}\in \mbf{T\s}
\end{alignat*}
due to
\begin{alignat*}{1}
\left(\os{\s}{\bnabla}\otimes \mbf{V},\mbf{n}\right)_{\mathbb{R}^3}
    &=\left(\mbf{g}^l\otimes\frac{\pl}{\pl x_\s^l}\left(V_i\mbf{g}^i+V^3\mbf{n}\right)(x_\s,t),\mbf{n}\right)_{\mathbb{R}^3}\\
    &=\left(\nabla_lV_i\mbf{g}^l\otimes\mbf{g}^i+V_ib^i_l\mbf{g}^l\otimes\mbf{n}
        +\frac{\pl V^3}{\pl x_\s^l}(x_\s,t)\mbf{g}^l\otimes\mbf{n}
        +V^3\mbf{g}^l\otimes\frac{\pl \mbf{n}}{\pl x_\s^l}(x_\s,t),\mbf{n}\right)_{\mathbb{R}^3}\\
    &=\left(V_ib^i_l+\frac{\pl V^3}{\pl x_\s^l}(x_\s,t)\right)\mbf{g}^l
     =\os{\s}{\bnabla}V^3+\mbf{V}\bcdot\mbf{K}\in\mbf{T\s}
\end{alignat*}

As a summary, it is attained
\begin{alignat*}{1}
\mbf{V}\otimes \bnabla=\left(\theta-\os{\s}{\bnabla}\bcdot\mbf{V}\right)\mbf{n}\otimes\mbf{n}
    +(\mbf{\omega}\times\mbf{n})\otimes \mbf{n}+(\mbf{W}\times\mbf{n})\otimes\mbf{n}
    -[[\mbf{V}\otimes\bnabla,\mbf{n}],\mbf{n}]
\end{alignat*}
To deal with the last term, one has
\begin{alignat*}{1}
    -[[\mbf{V}\otimes\bnabla,\mbf{n}],\mbf{n}]=-\left[ \left[\mbf{V}\otimes\os{\s}{\bnabla},\mbf{n}\right],\mbf{n}\right]
    =\mbf{V}\otimes\os{\s}{\bnabla}-\left( \mbf{V}\otimes\os{\s}{\bnabla},\mbf{n} \right)_{\mathbb{R}^3}\ots\mbf{n}
    =\mbf{V}\otimes\os{\s}{\bnabla}
\end{alignat*}

Consequently, one arrives at the identity that bridges the relation between the full dimensional velocity gradient and the one with respect to the surface gradient
\begin{alignat*}{1}
\mbf{V}\otimes \bnabla=\left(\theta-\os{\s}{\bnabla}\bcdot\mbf{V}\right)\mbf{n}\otimes\mbf{n}
    +(\mbf{\omega}\times\mbf{n})\otimes \mbf{n}+(\mbf{W}\times\mbf{n})\otimes\mbf{n}
    +\mbf{V}\otimes\os{\s}{\bnabla}
\end{alignat*}
Taking account of the conjugate identity, one ends the proof.

The proved identity bridges the relation between the whole strain and the one due purely to the deformation of the surface. Consequently, one has the relation
\begin{alignat*}{3}
    |\Delta\mbf{p}|_{\mathbb{R}^3}\os{\bdot}{\ol{|\Delta\mbf{p}|_{\mathbb{R}^3}}}
&  =\Delta\mbf{p}\bdot\mbf{D}\bdot\Delta\mbf{p}\\
&  =\left( \frac{\pl \mbf{V}}{\pl x^3}(x_\s,t)+\mbf{W}\ts\mbf{n}, \Delta\mbf{p} \right)_{\mathbb{R}^3}
    |\Delta\mbf{p}_\perp|_{\mathbb{R}^3}
   +\Delta\mbf{p}_\parallel\bdot\ \os{\Sigma}{\mbf{D}}\bdot\Delta\mbf{p}_\parallel\\
&  =\left( \frac{\pl \mbf{V}}{\pl x^3}(x_\s,t)+\mbf{W}\ts\mbf{n}, \Delta\mbf{p} \right)_{\mathbb{R}^3}
    |\Delta\mbf{p}_\perp|_{\mathbb{R}^3}
   +|\Delta\mbf{p}_\parallel|_{\mathbb{R}^3}\os{\bdot}{\ol{|\Delta\mbf{p}_\parallel|_{\mathbb{R}^3}}}
\end{alignat*}
where $\Delta\mbf{p}=\Delta\mbf{p}_\perp+\Delta\mbf{p}_\parallel$ denotes the directed line element,  $\Delta\mbf{p}_\perp \perp \mbf{T\Sigma}$ and $\Delta\mbf{p}_\parallel \in \mbf{T\Sigma}$ are directed line elements with respect to the normal direction and tangent space respectively. As indicated by \cite{XXL-2013}, $|\Delta\mbf{p}_\parallel|_{\mathbb{R}^3}\os{\bdot}{\ol{|\Delta\mbf{p}_\parallel|_{\mathbb{R}^3}}}
=\Delta\mbf{p}_\parallel\bdot\os{\Sigma}{\mbf{D}}\bdot\Delta\mbf{p}_\parallel$ is a kind of representations of the deformation of the boundary. In detail, it can be deduced
\begin{alignat*}{3}
& \frac{\pl\mbf{V}}{\pl x^3}(x_\s,t)
=\left( \frac{\pl V^j}{\pl x^3}(x_\s,t)-V^sb^j_s \right)\mbf{g}_j
+\frac{\pl V^3}{\pl x^3}(x_\s,t)\mbf{n}=:\frac{\pl\mbf{V}}{\pl \mbf{n}}(x_\s,t)\\
& \mbf{W}\ts\mbf{n}=\os{\Sigma}{\nb}V^3+V^sb^j_s\mbf{g}_j=\os{\Sigma}{\nb}V^3+\mbf{V}\bdot\mbf{K}
\in\mbf{T\Sigma}
\end{alignat*}
It is evident that $\frac{\pl\mbf{V}}{\pl \mbf{n}}(x_\s,t)$ represents the interaction between the boundary and fluid, $\mbf{W}\ts\mbf{n}$ is determined by the boundary.

\cite{WJZ-2005} deduced a kind of novel representation of the stain tensor on an arbitrary deformable boundary based on the triple decomposition of the velocity gradient and the intrinsic decomposition, that is
\begin{alignat*}{3}
    \mbf{D}
&  =\left( \theta-\os{\Sigma}{\bnb}\bdot\mbf{V} \right)\mbf{n}\otimes\mbf{n}
   +\frac{1}{2}(\mbf{\omega}\ts\mbf{n})\ots\mbf{n}+\frac{1}{2}\mbf{n}\ots(\mbf{\omega}\ts\mbf{n})\\
&  +[ (\mbf{W}\ts\mbf{n})\ots\mbf{n}+\mbf{n}\ots(\mbf{W}\ts\mbf{n}) ]
   -\os{Sym}{\ol{\left[\mbf{n},\left[\mbf{n},\mbf{V}\ots\os{\Sigma}{\bnb}\right]\right]}}
\end{alignat*}
The above representation can be deduced when the surface gradient of the velocity is rewritten as
\begin{alignat*}{3}
    \mbf{V}\ots\os{\Sigma}{\bnb}
&  =\mbf{n}\ots\left( \mbf{n}, \mbf{V}\ots\os{\Sigma}{\bnb}\right)_{\mathbb{R}^3}
   -\left[\mbf{n},\left[\mbf{n},\mbf{V}\ots\os{\Sigma}{\bnb}\right]\right]
   =\mbf{n}\ots(\mbf{W}\ts\mbf{n})-\left[\mbf{n},\left[\mbf{n},\mbf{V}\ots\os{\Sigma}{\bnb}\right]\right]
\end{alignat*}

It should be mentioned that as presented in this subsection the deduced representation of the strain tensor on an arbitrary deformable boundary with the adoption of the intrinsic decomposition is gained enlightenment from the work by \cite{WJZ-2005}.

\section{Levi-Civita gradient operator}

One can defines the so termed \emph{Levi-Civita connection operator} $\bnabla\equiv \boldsymbol{g}^l\nabla_\frac{\partial}{\partial x_\Sigma^l}$
\begin{alignat*}{3}
    \bnabla\circ-\mathbf{\Phi}
&   \equiv(\boldsymbol{g}^l\nabla_{\frac{\partial}{\partial x_\Sigma^l}})\circ-( \Phi^i_{\cdot j}\boldsymbol{g}_i\otimes\boldsymbol{g}^j +\Phi^i_{\cdot 3}\mbf{g}_i\otimes\mbf{n}+\Phi^3_{\cdot j}\mbf{n}\otimes\mbf{g}^j+\Phi^3_{\cdot 3}\mbf{n}\otimes\mbf{n})\\
& \triangleq\boldsymbol{g}^l\circ-\nabla_{\frac{\partial}{\partial x_\Sigma^l}}( \Phi^i_{\cdot j}\boldsymbol{g}_i\otimes\boldsymbol{g}^j +\Phi^i_{\cdot 3}\mbf{g}_i\otimes\mbf{n}+\Phi^3_{\cdot j}\mbf{n}\otimes\mbf{g}^j+\Phi^3_{\cdot 3}\mbf{n}\otimes\mbf{n})\\
&   =\nabla_l\Phi^i_{\cdot j}(\boldsymbol{g}^l\circ-\boldsymbol{g}_i)\otimes\boldsymbol{g}^j
    +\nb_l\Phi^i_{\cdot 3}(\boldsymbol{g}^l\circ-\boldsymbol{g}_i)\otimes\mbf{n}
    +\nb_l\Phi^3_{\cdot j}(\boldsymbol{g}^l\circ-\mbf{n})\otimes\mbf{g}^j\\
&   +\nb_l\Phi^3_{\cdot 3}(\mbf{g}^l\circ-\mbf{n})\otimes\mbf{n}
\end{alignat*}
As compared to the surface gradient tensor, Levi-Civita gradient tensor is just effective to the components/indeces of
the tensor corresponding to the tangent plane, i.e. $i$, $j$ in the above representations. In the present paper,
Levi-Civita gradient operator takes the same denotation $\bnb$ as the full dimensional gradient operator since the
concrete meaning of $\bnb$ can be determined in certain relations.

\subsection{Some primary identities in vorticity dynamics of two dimensional flows on fixed smooth surfaces}

\begin{proposition}
The Levi-civita gradient operator with respect to initial physical configuration, denoted by  $\overset{\circ}{\bnabla}\triangleq\boldsymbol{G}^L\overset{\circ}{\nabla}_{\frac{\partial}{\partial\xi^L_\s}}$, can be similarly defined. And the order between $\overset{\circ}{\bnabla}$ and material derivative can be changed, namely
\begin{equation*}
    \os{\circ}{\bnb}\circ-{\dot{\mbf{\Phi}}}=\os{\bdot}{\overline{\os{\circ}{\bnb}\circ-\mbf{\Phi}}},\quad
    \forall\,\mbf{\Phi}\in\mathscr{T}^p(\mathbb{R}^{m+1})
\end{equation*}
\end{proposition}

\noindent\textbf{Proof}\quad
One has, say $\mbf{\Phi}\in\mathscr{T}^2(\mathbb{R}^{3})$
\begin{alignat*}{3}
    \os{\circ}{\bnb}\circ-\mbf{\Phi}
&   \equiv(\boldsymbol{G}^L\os{\circ}{\nb}_{\frac{\partial}{\partial \xi^L_\Sigma}})\circ-( \Phi^A_{\cdot B}\boldsymbol{G}_A\otimes\boldsymbol{G}^B +\Phi^A_{\cdot 3}\mbf{G}_A\otimes\mbf{N}+\Phi^3_{\cdot B}\mbf{N}\otimes\mbf{G}^B+\Phi^3_{\cdot 3}\mbf{N}\otimes\mbf{N})\\
&   =\os{\circ}{\nb}_L\Phi^A_{\cdot B}(\boldsymbol{G}^L\circ-\boldsymbol{G}_A)\otimes\boldsymbol{G}^B
    +\os{\circ}{\nb}_L\Phi^A_{\cdot 3}(\boldsymbol{G}^L\circ-\boldsymbol{G}_A)\otimes\mbf{N}
    +\os{\circ}{\nb}_L\Phi^3_{\cdot B}(\boldsymbol{G}^L\circ-\mbf{N})\otimes\mbf{G}^B\\
&   +\os{\circ}{\nb}_L\Phi^3_{\cdot 3}(\mbf{G}^L\circ-\mbf{N})\otimes\mbf{N}
\end{alignat*}
where $\{\mbf{G}_A(\xi_\Sigma)\}^2_{A=1}$ and $\{\mbf{G}^A(\xi_\Sigma)\}^2_{A=1}$ denote the covariant and contra-variant bases with respect to the initial physical configuration, and $\mbf{N}(\xi_\Sigma)$ is the corresponding normal vector. All of them are independent on the time. Thanks to the relationships
\begin{alignat*}{3}
&  \frac{\pl}{\pl t}\left(\os{\circ}{\nb}_L\Phi^A_{\cdot B}\right)(\xi_\Sigma,t)
   =\os{\circ}{\nb}_L\left( \frac{\pl \Phi^A_{\cdot B} }{\pl t}(\xi_\Sigma,t) \right)\\
&  \frac{\pl}{\pl t}\left(\os{\circ}{\nb}_L\Phi^A_{\cdot 3}\right)(\xi_\Sigma,t)
   =\os{\circ}{\nb}_L\left( \frac{\pl \Phi^A_{\cdot 3} }{\pl t}(\xi_\Sigma,t) \right);\quad
   \frac{\pl}{\pl t}\left(\os{\circ}{\nb}_L\Phi^3_{\cdot B}\right)(\xi_\Sigma,t)
   =\os{\circ}{\nb}_L\left( \frac{\pl \Phi^3_{\cdot B} }{\pl t}(\xi_\Sigma,t) \right)\\
&  \frac{\pl}{\pl t}\left(\os{\circ}{\nb}_L\Phi^3_{\cdot 3}\right)(\xi_\Sigma,t)
   =\os{\circ}{\nb}_L\left( \frac{\pl \Phi^3_{\cdot 3} }{\pl t}(\xi_\Sigma,t) \right)
\end{alignat*}
the proof is completed.

Firstly, the following identity has been derived
\begin{proposition}
\begin{equation*}
    \left[\overset{\circ}{\bnabla}\times(\boldsymbol{b}\bcdot\mathbf{F})\right]\bcdot\boldsymbol{N}
   =|\mathbf{F}|\left(\bnabla\times\boldsymbol{b} \right)\bcdot\boldsymbol{n},\quad
    \forall\,\boldsymbol{b}\in\mathbb{R}^3,
    \quad |\mathbf{F}|:=\frac{\sqrt{g}}{\sqrt{G}}\det\left[\frac{\partial x^i}{\partial\xi^A}\right](\xi,t)
\end{equation*}
where $\overset{\circ}{\bnabla}\triangleq\boldsymbol{G}^L\overset{\circ}{\nabla}_{\frac{\partial}{\partial\xi^L_\s}}$ and $\bnabla\triangleq\boldsymbol{g}^l\nabla_{\frac{\partial}{\partial x^l_\s}}$ are \emph{\textit{Levi-Civita connection operators}}, $\boldsymbol{N}$ and $\boldsymbol{n}$ are surface normal vectors corresponding to the initial and current physical configurations respectively, $\sqrt{G}:=[\boldsymbol{G}_1,\boldsymbol{G}_2,\boldsymbol{N}]$,
$\sqrt{g}:=[\boldsymbol{g}_1,\boldsymbol{g}_2,\boldsymbol{n}]$.
\end{proposition}

\noindent\textbf{Proof}\quad
\begin{eqnarray}
  && \left[\overset{\circ}{\bnabla}\times(\boldsymbol{b}\bcdot\mathbf{F})\right]\bcdot\boldsymbol{N}
    =\left[ \left(\boldsymbol{G}^B\overset{\circ}{\nabla}_{\frac{\partial}{\partial\xi_\Sigma^B}}\right)\times\left(
     b_i\frac{\partial x_\Sigma^i}{\partial \xi_\Sigma^A}(\xi_\Sigma,t)\boldsymbol{G}^A \right)  \right]\bcdot\boldsymbol{N}
    =\overset{\circ}{\nabla}_B\left(b_i\frac{\partial x_\Sigma^i}{\partial\xi_\Sigma^A}(\xi_\Sigma,t)\epsilon^{BA3} \right)
    \nonumber\\
  &&=\eps^{BA3}\overset{\circ}{\nabla}_B\left(b_i\frac{\partial x_\Sigma^i}{\partial\xi_\Sigma^A}(\xi_\Sigma,t)\right)
    =\eps^{BA3}\left[ \frac{\partial}{\partial\xi_\Sigma^B}\left(b_i\frac{\partial x_\Sigma^i}{\partial\xi_\Sigma^A} \right)(\xi_\Sigma,t)
    -\Gamma^L_{BA}\left(b_i\frac{\partial x_\Sigma^i}{\partial\xi_\Sigma^L}(\xi_\Sigma,t) \right) \right]
    \nonumber\\
  &&=\eps^{BA3} \frac{\partial}{\partial\xi_\Sigma^B}\left(b_i\frac{\partial x_\Sigma^i}{\partial\xi_\Sigma^A} \right)(\xi_\Sigma,t)
    =\frac{\partial b_i}{\partial x^s_\Sigma}(x_\Sigma,t)\left[ \eps^{BA3}\frac{\partial x^s_\Sigma}{\partial\xi^B_\Sigma}(\xi_\Sigma,t)
    \frac{\partial x^i_\Sigma}{\partial\xi_\Sigma^A}(\xi_\Sigma,t) \right]
    \nonumber\\
  &&=\frac{1}{\sqrt{G}}\det\left[\frac{\partial x_\Sigma^i}{\partial\xi_\Sigma^A}\right](\xi_\Sigma,t)
     e^{si3}\frac{\partial b_i}{\partial x_\Sigma^s}(x_\Sigma,t)
    =\frac{\sqrt{g}}{\sqrt{G}}\det\left[\frac{\partial x_\Sigma^i}{\partial\xi_\Sigma^A}\right](\xi_\Sigma,t)(\eps^{si3}\nabla_sb_i)
    \nonumber\\
  &&=|\mathbf{F}|(\eps^{si3}\nabla_sb_i)
  \nonumber
\end{eqnarray}

On the other hand, one has
\begin{alignat*}{3}
    \bnb\times\mbf{b}
&  =\bnb\times(b^i\mbf{g}_i+b^3\mbf{n})=\mbf{g}^l\times\nb_{\frac{\pl}{\pl x^l_\Sigma}}(b^i\mbf{g}_i+b^3\mbf{n})
   =\mbf{g}^l\times\left( \nb_l b_i \mbf{g}^i+\frac{\pl b^3}{\pl x^l_\Sigma}\mbf{n} \right)\\
&  =\eps^{li3} \nb_l b_i\mbf{n}+ \eps^{l3k}\frac{\pl b^3}{\pl x^l_\Sigma}\mbf{g}_k
\end{alignat*}
This ends the proof.

As an application, one has
\begin{equation*}
    \omega^3 :=(\bnabla\times\boldsymbol{V})\bcdot\boldsymbol{n}=\frac{1}{|\mathbf{F}|}
    \left[ \overset{\circ}{\bnabla}\times(\boldsymbol{V}\bcdot\mathbf{F}) \right]\bcdot\boldsymbol{N}
\end{equation*}
Subsequently, the governing equation of vorticity can be deduced
\begin{eqnarray}
    \dot{\omega}^3
 &&=-\frac{\theta}{|\mathbf{F}|}
    \left[ \overset{\circ}{\bnabla}\times(\boldsymbol{V}\bcdot\mathbf{F}) \right]\bcdot\boldsymbol{N}
   +\frac{1}{|\mathbf{F}|}
    \left[ \overset{\circ}{\bnabla}\times(\boldsymbol{a}\bcdot\mathbf{F})
          +\overset{\circ}{\bnabla}\times\left(\, \boldsymbol{V}
          \bcdot \left( \boldsymbol{V}\otimes\overset{\Sigma}{\bnabla} \right)\bcdot\mathbf{F} \,\right)
    \right]\bcdot\boldsymbol{N}
    \nonumber\\
 &&=-\theta(\bnabla\times\boldsymbol{V})\bcdot\boldsymbol{n}+(\bnabla\times\boldsymbol{a})\bcdot\boldsymbol{n}
   +\left( \bnabla\times\bnabla\left(\frac{|\boldsymbol{V}|^2}{2}\right) \right)\bcdot\boldsymbol{n}
   =-\theta\omega^3+(\bnabla\times\boldsymbol{a})\bcdot\boldsymbol{n}
    \nonumber
\end{eqnarray}
where the identities $\dot{\mathbf{F}}=\left(\boldsymbol{V}\otimes\overset{\Sigma}{\bnabla}\right)\bcdot\mathbf{F}$ and $\mathrm{d}|\mathbf{F}|/\mathrm{d}t=\theta|\mathbf{F}|$ are utilized.

\begin{proposition}
\begin{equation}
    \bnabla\times(\bnabla\times\boldsymbol{b})=\bnabla(\bnabla\bcdot\boldsymbol{b})
    -\boldsymbol{\Delta}\boldsymbol{b}+K_G\boldsymbol{b},\quad
    \forall\,\boldsymbol{b}\in\boldsymbol{T\Sigma},
    \quad\boldsymbol{\Delta}\boldsymbol{b}\triangleq\bnabla\bcdot(\bnabla\otimes\boldsymbol{b})
    \nonumber
\end{equation}
\end{proposition}
\noindent\textbf{Proof}\quad
\begin{alignat*}{3}
    \bnabla\times(\bnabla\times\boldsymbol{b})
   &=\bnabla\times\left[ \left(\boldsymbol{g}^p\nabla_{\frac{\partial}{\partial x^p_\s}}\right) \times
     \left( b_i\boldsymbol{g}^i\right) \right]
    =\left( \boldsymbol{g}^q\nabla_{\frac{\partial}{\partial x^q_\s}} \right) \times \left[(\nabla_p b_i)\epsilon^{pi3}\boldsymbol{n} \right]\\
   &=\eps^{3kq}\eps_{3pi}\nabla_q(\nabla^p b^i)\boldsymbol{g}_k
    =(\delta^k_p\delta^q_i-\delta^q_p\delta^k_i)\nabla_q(\nabla^p b^i)\boldsymbol{g}_k
    =\nabla_i(\nabla^k b^i)\boldsymbol{g}_k-\nabla_p(\nabla^p b^i)\boldsymbol{g}_i\\
   &=\nabla_i(\nabla^k b^i)\boldsymbol{g}_k-\boldsymbol{\Delta}\boldsymbol{b}
\end{alignat*}
Furthermore, one has
\begin{alignat*}{3}
    \nabla_i(\nabla^k b^i)\boldsymbol{g}_k
  &=[\nabla^k(\nabla_i b^i)+R^{i\cdot\cdot k}_{\cdot si\cdot} b^s]\boldsymbol{g}_k
   =\bnabla(\bnabla\cdot\boldsymbol{b}) + K_G(\delta^i_i\delta^k_s-g_{si}g^{ik})b^s\boldsymbol{g}_k \\
  &=\bnabla(\bnabla\cdot\boldsymbol{b}) + K_G\boldsymbol{b}
\end{alignat*}
It is the end of the proof.

The well known Stokes-Helmholtz decomposition in the present case as shown below takes the different form as compared to the one for Euclid space.
\begin{proposition}[Stokes-Helmholtz Decomposition]
For any $\mbf{b}\in\mbf{T\Sigma}$, one has $\mbf{b}=\bnb\phi+\bnb\times(\psi\mbf{n})$, where $\phi$ and $\psi$ can be termed as the tangent plane potential and the normal potential respectively. Both of them are determined by the Possion equations
\begin{alignat*}{3}
&   \Delta\phi:=\bnb\bdot(\bnb\phi)=g^{ij}\left[\frac{\pl^2\phi}{\pl x^i_\Sigma\pl x^j_\Sigma}(x_\Sigma,t)
-\Gamma^k_{ij}\frac{\pl\phi}{\pl x^k_\Sigma}(x_\Sigma,t)\right]=\bnb\bdot\mbf{b}\\
&   \Delta\psi:=\bnb\bdot(\bnb\psi)=g^{ij}\left[\frac{\pl^2\psi}{\pl x^i_\Sigma\pl x^j_\Sigma}(x_\Sigma,t)
-\Gamma^k_{ij}\frac{\pl\psi}{\pl x^k_\Sigma}(x_\Sigma,t)\right]=-(\bnb\times\mbf{b})\bdot\mbf{n}
\end{alignat*}
\end{proposition}

As an application, let us focus on two dimensional flows on general fixed smooth surface. The velocity $\mbf{V}\in\mbf{T\Sigma}$ can be represented as
\begin{equation*}
    \mbf{V}=\bnb\phi+\bnb\times(\psi\mbf{n}),\quad\mbox{with}\,
    \left\{
    \begin{array}{l}
    \Delta\phi=\bnb\bdot\mbf{V}=:\theta\\
    \Delta\psi=-(\bnb\times\mbf{V})\bdot\mbf{n}=-\omega^3
    \end{array}
    \right.
\end{equation*}
where $\theta$ is termed as the dilation quantity, the vorticity is defined as $\mbf{\omega}\triangleq\omega^3\mbf{n}=\bnb\times\mbf{V}$. In the case of $\theta=0$ i.e. the flow is incompressible, the velocity takes the representation $\mbf{V}=\bnb\ts(\psi\mbf{n})$ namely $V^i=-\eps^{3ji}\frac{\pl\psi}{\pl x^j_\s}(x_\s,t)$, in which $\psi$ is generally termed as stream function. Based on the intrinsic generalized Stokes formula of the first kind, one has
\begin{equation*}
    \oint_{\pl\Sigma}\,(\mbf{\tau}\ts\mbf{n})\bdot\mbf{V}\,dl
   =\int_\Sigma\,\os{\Sigma}{\bnb}\bdot\mbf{V}\,d\sigma
   =\int_\Sigma\,\bnb\bdot\mbf{V}\,d\sigma=0
\end{equation*}
In other words, the integral of $\int_{\pl\Sigma}\,(\mbf{\tau}\ts\mbf{n})\bdot\mbf{V}\,dl$ is independent on the integral paths. Furthermore, the stream function can be determined through
\begin{equation*}
    \psi(\mbf{r},t)=\psi(\mbf{r}_0,t)+\int_{\pl\Sigma}\,(\mbf{\tau}\ts\mbf{n})\bdot\mbf{V}\,dl
\end{equation*}
where $\mbf{r}_0$ represents the reference point that can be an arbitrary point on the surface.

We have attained the governing equation of mass conservation
\begin{equation*}
    \dot{\rho}+\rho\theta=\left[ \frac{\pl\rho}{\pl t}(x_\s,t)+\mbf{V}\bdot(\bnb\rho) \right] +\rho\theta=0
\end{equation*}
where $\rho$ is the surface density.

Generally, denoted by $\mbf{t}$ the surface stress that satisfies $\rho\mbf{a}=\bnb\bdot\mbf{t}+\rho\mbf{f}_\s\in\mathbb{R}^3$ can be assumed that
\begin{equation*}
    \mbf{t}=(\gamma-p)\mbf{I}+\mu\left( \mbf{V}\otimes\bnb+\bnb\otimes\mbf{V} \right),
    \quad\mbf{I}=\delta^i_j\mbf{g}_i\otimes\mbf{g}^j
\end{equation*}
where $\gamma$ and $\mu$ denote coefficients of surface tension and inner fraction/viscousity respectively, $p$ is the inner pressure.

Subsequently, the governing equations of momentum conservation with respect to the tangent plane and normal direction can be deduced
\begin{alignat*}{3}
&    \rho\left[ \frac{\pl\mbf{V}}{\pl t}(x_\s,t)+\mbf{V}\bdot(\bnb\otimes\mbf{V})\right]
   =-\bnb p+\mu\left( \Delta\mbf{V}+\bnb\theta+K_G\mbf{V} \right)+\rho\mbf{f}_{sur,\s}\,\in\mbf{T\Sigma}\\
&    \rho(\mbf{V}\otimes\mbf{V}):\mbf{K}=H(\gamma-p)+2\mu(\mbf{V}\otimes\bnb):\mbf{K}+\rho f_{sur,n}\,\in\mathbb{R}
\end{alignat*}
where $\mbf{f}_{sur,\s}$ and $f_{sur,n}$ denote the densities of the surface forces on the tangent plane and normal direction respectively.

Taking $\bnb\bdot$ on both sides of the equation with respect to the tangent plane, one arrives at the governing equation of the dilation
\begin{alignat*}{3}
    \dot{\theta}=
&  -\left[ (\mbf{V}\otimes\bnb):(\bnb\otimes\mbf{V})+K_G|\mbf{V}|^2 \right]
   +\left[ \frac{1}{\rho^2}\bnb\rho\bdot\bnb p-\frac{1}{\rho}\Delta p \right]\\
&  -\frac{\mu}{\rho^2}\bnb\rho\bdot(\Delta\mbf{V}+\bnb\theta+K_G\mbf{V})
   +\frac{2\mu}{\rho}\left[ \Delta\theta+\bnb\bdot(K_G\mbf{V}) \right]+\bnb\bdot\mbf{f}_{sur}\\
  =& -\left[ \mbf{D}:\mbf{D}-\frac{|\mbf{\omega}|^2}{2}  +K_G|\mbf{V}|^2 \right]
   +\left[ \frac{1}{\rho^2}\bnb\rho\bdot\bnb p-\frac{1}{\rho}\Delta p \right]\\
&  -\frac{\mu}{\rho^2}\bnb\rho\bdot\left[ -\bnb\times\mbf{\omega}+2( \bnb\theta+K_G\mbf{V} ) \right]
   +\frac{2\mu}{\rho}\left[ \Delta\theta+\bnb\bdot(K_G\mbf{V}) \right]+\bnb\bdot\mbf{f}_{sur}
\end{alignat*}
where denoted by $\mbf{D}$ the strain tensor for two dimensional flows on fixed surfaces is defined by $\frac{1}{2}(\mbf{V}\ots\bnb+\bnb\ots\mbf{V})$.

On the other hand, the governing equation of the vorticity takes the following form
\begin{alignat*}{3}
    \dot{\omega}^3
&   =-\theta\omega^3
   -\frac{1}{\rho^2}[\bnb\rho,-\bnb p+\mu[-\bnb\times\mbf{\omega}+2\,(\bnb\theta+K_G\mbf{V})],\mbf{n}]\\
&   +\frac{\mu}{\rho}[\Delta\mbf{\omega}+2\bnb\times(K_G\mbf{V})]\bdot\mbf{n}
   -\frac{1}{\rho^2}[\bnb\rho,\mbf{f}_{sur},\mbf{n}]+\frac{1}{\rho}(\bnb\times\mbf{f}_{sur})\bdot\mbf{n}
\end{alignat*}

As a summary, one can simulate generally spatial-temporal evolutions of compressible two dimensional flows on general fixed smooth surfaces through the governing equations of dilation and vorticity accompanying with the Possion equations for the velocity potentials, in addition the distribution of inner pressure can be updated based on the equation of momentum conservation in the normal direction.

\subsection{Some identities of affine surface tensors}

\begin{proposition}
\begin{equation*}
    \bnb\times\mbf{\Phi}\times\bnb
   =\left[ \bnb\bdot\mbf{\Phi}\bdot\bnb-\mbf{\Delta}(tr\mbf{\Phi}) \right]\mbf{n}\otimes\mbf{n},
   \quad\forall\,\mbf{\Phi}\in\mathscr{T}^2(\mbf{T\Sigma})
\end{equation*}
\end{proposition}
\noindent\textbf{Proof}\quad

Firstly, one should show that
\begin{equation*}
    (\bnb\times\mbf{\Phi})\times\bnb=\bnb\times(\mbf{\Phi}\times\bnb)=:\bnb\times\mbf{\Phi}\times\bnb
\end{equation*}
The left hand side can be calculated as follows
\begin{alignat*}{3}
    (\bnb\ts\mbf{\Phi})\ts\bnb
&  =\left[ \left(\mbf{g}^p\nb_{\frac{\pl}{\pl x^p_\s}}\right)\ts\left(\Phi_{ij}\mbf{g}^i\ots\mbf{g}^j\right) \right]
    \ts\bnb
   =\left[(\nb_p\Phi_{ij})\eps^{pi3}\mbf{n}\ots\mbf{g}^j\right]\ts\left(\mbf{g}^q\nb_{\frac{\pl}{\pl x^q_\s}}\right)\\
&  =(\nb_q\nb^p\Phi^i_{\cdot j})\eps_{pis}\eps^{jq3}\mbf{n}\ots\mbf{n}
   =(\delta^j_p\delta^q_i-\delta^j_i\delta^q_p)(\nb_q\nb^p\Phi^i_{\cdot j})\mbf{n}\ots\mbf{n}\\
&  =\left[\nb_i\nb^j\Phi^i_{\cdot j}-\nb_q\nb^q\Phi^i_{\cdot i}\right]\mbf{n}\ots\mbf{n}
   =\left[\nb_i\nb^j\Phi^i_{\cdot j}-\mbf{\Delta}(tr\Phi)\right]\mbf{n}\ots\mbf{n}
\end{alignat*}
where
\begin{alignat*}{3}
    \nb_i\nb^j\Phi^i_{\cdot j}
&  =\nb^j\nb_i\Phi^i_{\cdot j}+R^{i\cdot\cdot j}_{\cdot si\cdot}\Phi^s_{\cdot j}
                              +R^{\cdot s\cdot j}_{j\cdot i\cdot}\Phi^i_{\cdot s}\\
&  =\nb^j\nb_i\Phi^i_{\cdot j}+K_G\left(\delta^i_i\delta^j_s-g_{si}g^{ij}\right)\Phi^s_{\cdot j}
                              +K_G\left(g_{ji}g^{sj}-\delta^s_i\delta^j_j\right)\Phi^i_{\cdot s}\\
&  =\nb^j\nb_i\Phi^i_{\cdot j}+K_G\Phi^j_{\cdot j}-K_G\Phi^i_{\cdot i}=\nb^j\nb_i\Phi^i_{\cdot j}
   =\bnb\cdot\mbf{\Phi}\cdot\bnb
\end{alignat*}
As a summary, one arrives at
\begin{equation*}
    (\bnb\ts\mbf{\Phi})\ts\bnb=\left[ \bnb\cdot\mbf{\Phi}\cdot\bnb-\mbf{\Delta}(tr\Phi) \right]\mbf{n}\ots\mbf{n}
\end{equation*}
On the other hand, $\bnb\times(\mbf{\Phi}\times\bnb)$ can similarly calculated with the same representation. This ends the proof.

Subsequently, it is evident that
\begin{equation*}
    \bnb\times\mbf{\Phi}\times\bnb=\mbf{0}  \quad\Leftrightarrow\quad
    \bnb\bdot\mbf{\Phi}\bdot\bnb-\mbf{\Delta}(tr\mbf{\Phi})=0\in\mathbb{R}
\end{equation*}

\begin{proposition}
\begin{equation*}
    \bnb\times(\bnb\times\mbf{\Phi})
   =\bnb\ots(\bnb\bdot\mbf{\Phi})-\Delta\mbf{\Phi}+K_G(\mbf{\Phi}+\mbf{\Phi}^*)-K_G(tr\mbf{\Phi})\mbf{I},
   \quad\forall\,\mbf{\Phi}\in\mathscr{T}^2(\mbf{T\Sigma})
\end{equation*}
\end{proposition}

\noindent\textbf{Proof}\quad

On the left hand side, one can do the following deduction
\begin{alignat*}{3}
    \bnb\ts\bnb\ts\mbf{\Phi}
&  =\bnb\ts\left[ \left(\mbf{g}^p\nb_{\frac{\pl}{\pl x^p_\s}}\right)\ts\left(\Phi_{ij}\mbf{g}^i\ots\mbf{g}^j\right) \right]
   =\left(\mbf{g}^q\nb_{\frac{\pl}{\pl x^q_\s}}\right)\ts\left[\nb_p\Phi_{ij}\eps^{pis}\mbf{n}\ots\mbf{g}^j\right]\\
&  =(\nb_q\nb_p\Phi_{ij})\eps^{q3k}\eps^{pi3}\mbf{g}_k\ots\mbf{g}^j
   =(\nb^q\nb_p\Phi_{ij})\eps_{3kq}\eps^{pi3}\mbf{g}^k\ots\mbf{g}^j\\
&  =(\nb^q\nb_p\Phi_{ij})(\delta^p_k\delta^i_q-\delta^p_q\delta^i_k)\mbf{g}^k\ots\mbf{g}^j
   =\nb^i\nb_k\Phi_{ij}\mbf{g}^k\ots\mbf{g}^j- \nb^p\nb_p\Phi_{ij}\mbf{g}^i\ots\mbf{g}^j\\
&  =\nb^i\nb_k\Phi_{ij}\mbf{g}^k\ots\mbf{g}^j- \mbf{\Delta}\mbf{\Phi}
\end{alignat*}
where the first term on the right hand side can be processed as follows
\begin{alignat*}{3}
&   \nb^i\nb_k\Phi_{ij}\,\mbf{g}^k\ots\mbf{g}^j
   =\left[ \nb_k\nb^i\Phi_{ij}+R^{\cdot si\cdot}_{i\cdot\cdot k}\Phi_{sj}
                              +R^{\cdot si\cdot}_{j\cdot\cdot k}\Phi_{is} \right]\mbf{g}^k\ots\mbf{g}^j\\
&  =(\nb_k\nb^i\Phi_{ij})\mbf{g}^k\ots\mbf{g}^j
    +K_G\left(\delta^i_i\delta^s_k-g^{si}g_{ik}\right)\Phi_{sj}\mbf{g}^k\ots\mbf{g}^j
    +K_G\left(\delta^i_j\delta^s_k-g^{si}g_{jk}\right)\Phi_{is}\mbf{g}^k\ots\mbf{g}^j\\
&  =\bnb\ots(\bnb\bdot\mbf{\Phi})+K_G(\mbf{\Phi}+\mbf{\Phi}^*)-K_G(tr\mbf{\Phi})\mbf{I}
\end{alignat*}
It is the end of the proof.

As indicated in this subsection, the change of the order of covariant or contra-variant differential operators defined on surfaces must be related to Riemann-Christoffel tensor. Consequently, the action of Levi-Civita gradient operators more than twice will generally lead to the appearance of the curvatures, such as Laplacian operator $\mbf{\Delta}\mbf{\Phi}\triangleq \bnb\bdot(\bnb\ots\mbf{\Phi})$, double curl operator $\bnb\ts(\bnb\ts\mbf{\Phi})$  and so on.

\section{Conclusion Remarks}

Two kinds of differential operators on the surface have been studied including definitions, properties and applications.

The first kind of differential operators on the surface is termed as the surface gradient operator. It can be taken as
the derivative of a tensor field defined on the surface. Consequently, the partial derivative of a tensor field with
respect to a curvilinear coordinate of the surface can be determined. According to the differential calculus in normed
linear tensor spaces, the order of partial derivatives of a tensor field can be exchanged as the regularity of the
tensor filed is provided. Firstly, all kinds of the intrinsic generalized Stokes formulas have been derived in which
the relation between the full dimensional gradient operator and the surface gradient one is adopted. Particularly, the
intrinsic generalized Stokes formula of the second kind have been used to deduce some integral identities studies by
\cite{Yin-2008a}; to deduce the governing differential equations of momentum and moment of momentum conservations for continuous mediums whose geometrical configurations can be taken as surfaces that cover the governing equations for thin enough plates and shells as deduced by \cite{Chien-1941}; to deduce a differential identity that plays the essential
role to attain the representation of the vorticity flux on an arbitrary deformable solid boundary. Secondly,
the primary properties of the deformation gradient tensor with respect to the continuous mediums whose geometrical
configurations are surfaces with any finite dimensionality have been deduced. The deduction mistake made by
\cite{Aris-1962} in his studies on two dimensional flows on an arbitrary fixed surface have been pointed out in
detail. Thirdly, the representations of strain tensor on an arbitrary deformable surface have been studied. The
original result was attained by \cite{WJZ-2005}.

The second kind of differential operators on the surface is termed as
Levi-Civita gradient operator. Its definition is based on general Levi-Civita connection on the surface that is a
typical Riemann manifold. It has been indicated that Levi-Civita operator is just effective to the indices with respect
to the tangent space. Figuratively, Levi-Civita gradient of a tensor field just represents the feeling of an observer
standing on the surface who is insensitive to the change in the normal direction, however the surface gradient
reflects the whole change of a tensor field as viewed from the outer space of the surface. The essential property of
Levi-Civita gradient operator is that the change of the order of covariant or contra-variant
derivatives/differentiations should be related to Riemann-Christoffel tensor. Some differential identities have been
set up based on Levi-Civita gradient operator that constitute the foundation of the theoretical framework of vorticity
dynamics for two dimensional flows on an arbitrary fixed surface as indicated by \cite{XXL-2013-arXiv}. In the present case, there are some additional terms in the governing equations of momentum and vorticity that include curvatures of the surface explicitly. It implies that the geometrical quantities accompanying with the mechanical ones may appear explicitly in the governing equations of nature laws as the physical configurations of continuous mediums can be taken as surfaces.

\section*{Acknowledgements}

This work is supported by National Nature Science Foundation of China (Grant No.$11172069$) and  some key projects of
education reforms issued by the Shanghai Municipal Education Commission ($2011$).

\ \\

\end{document}